\documentclass[12pt]{article}
\usepackage{axodraw,a4wide,epsfig}
\usepackage{scalefnt}

\newcommand{\nn}{\nonumber}
\newcommand{\be}{\begin{equation}}
\newcommand{\ee}{\end{equation}}
\newcommand{\bea}{\begin{eqnarray}}
\newcommand{\eea}{\end{eqnarray}}
\newcommand{\bfm}[1]{\mbox{\boldmath$#1$}}

\catcode`@=11
\newcount\@tempcntc
\def\@citex[#1]#2{\if@filesw\immediate\write\@auxout{\string\citation{#2}}\fi
  \@tempcnta\z@\@tempcntb\m@ne\def\@citea{}\@cite{\@for\@citeb:=#2\do
    {\@ifundefined
       {b@\@citeb}{\@citeo\@tempcntb\m@ne\@citea\def\@citea{,}{\bf ?}\@warning
       {Citation `\@citeb' on page \thepage \space undefined}}%
    {\setbox\z@\hbox{\global\@tempcntc0\csname b@\@citeb\endcsname\relax}%
     \ifnum\@tempcntc=\z@ \@citeo\@tempcntb\m@ne
       \@citea\def\@citea{,}\hbox{\csname b@\@citeb\endcsname}%
     \else
      \advance\@tempcntb\@ne
      \ifnum\@tempcntb=\@tempcntc
      \else\advance\@tempcntb\m@ne\@citeo
      \@tempcnta\@tempcntc\@tempcntb\@tempcntc\fi\fi}}\@citeo}{#1}}
\def\@citeo{\ifnum\@tempcnta>\@tempcntb\else\@citea\def\@citea{,}%
  \ifnum\@tempcnta=\@tempcntb\the\@tempcnta\else
   {\advance\@tempcnta\@ne\ifnum\@tempcnta=\@tempcntb \else \def\@citea{--}\fi
    \advance\@tempcnta\m@ne\the\@tempcnta\@citea\the\@tempcntb}\fi\fi}
\catcode`@=12

\begin{document}

\title{\vskip-3cm{\baselineskip14pt
\centerline{\normalsize DESY 04-099\hfill}
\centerline{\normalsize TTP04-10\hfill}
\centerline{\normalsize UB-ECM-PF-04-16 \hfill}
\centerline{\normalsize  June 2004\hfill}
}
\vskip1.5cm
Spin Dependence of Heavy Quarkonium Production and Annihilation Rates:
\\
Complete Next-to-Next-to-Leading Logarithmic result}
\author{A.A. Penin$^{a,b}$,
A. Pineda$^c$, V.A. Smirnov$^{d,e}$, M. Steinhauser$^e$\\[0.5cm]
{\small\it $^a$ Institut f{\"u}r Theoretische Teilchenphysik,Universit{\"a}t Karlsruhe,}\\
{\small\it  76128 Karlsruhe, Germany}\\
{\small\it $^b$ Institute for Nuclear Research, Russian Academy of Sciences,}\\
{\small\it 60th October Anniversary Prospect 7a, 117312 Moscow, Russia}\\
{\small\it $^c$ Dept. d'Estructura i Constituents de la Mat\`eria, U. Barcelona,}\\
{\small\it Diagonal 647, E-08028 Barcelona, Catalonia,  Spain}\\
{\small\it $^d$ Institute for Nuclear Physics, Moscow State University,}\\
{\small\it 119992 Moscow, Russia}\\
{\small\it $^e$ II. Institut f\"ur Theoretische Physik, Universit\"at Hamburg,}\\
{\small\it Luruper Chaussee 149, 22761 Hamburg, Germany}}\date{}

\maketitle

\thispagestyle{empty}

\begin{abstract}
The ratio of the photon mediated production or annihilation rates of
spin triplet and spin singlet heavy quarkonium states is computed to the
next-to-next-to-leading logarithmic accuracy within the nonrelativistic
renormalization group approach.  The result is presented in analytical
form and applied to the phenomenology of $t\bar{t}$, $b\bar{b}$ and
$c\bar{c}$ systems. The use of the nonrelativistic renormalization group
considerably improves the behaviour of the perturbative expansion and is
crucial for accurate theoretical analysis. For bottomonium decays we
predict $\Gamma(\eta_b(1S) \rightarrow \gamma\gamma)=0.659\pm 0.089
({\rm th.})  {}^{+0.019}_{-0.018} (\delta \alpha_{\rm s})\pm 0.015 ({\rm
exp.})\; {\rm keV}$.  Our results question the accuracy of the
existing extractions of the strong coupling constant from the bottomonium
annihilation. As a by-product we obtain a novel result for the ratio of the
ortho- and parapositronium decay rates: the corrections of order
$\alpha^4\ln^2\alpha$ and $\alpha^5\ln^3\alpha$.
\\[2mm]
\end{abstract}

\newpage


\section{Introduction}
\label{int}

Annihilation of a heavy quarkonium into leptons, photons or light
hadrons, as well as its production in $e^+e^-$ or $\gamma\gamma$
collisions, has been a subject of numerous investigations starting from
earliest applications of perturbative quantum chromodynamics (QCD)
\cite{AppPol} and has by now become a classical problem. The
annihilation of the nonrelativistic bound state possesses a highly
sophisticated multi-scale dynamics and, not surprisingly, was the
original testing ground for the effective field theory of nonrelativistic
QCD (NRQCD) \cite{CasLep,BBL}.  On the phenomenological side, the ratio
of the bottomonium annihilation rates into leptons and light hadrons is
currently used to determine the value of the strong coupling constant
$\alpha_s$ \cite{Hag}.  Furthermore, the study of $t\bar t$ threshold
production at a future Linear Collider could even allow us to probe
Higgs-boson-induced effects \cite{MarMiq}.  The calculation of the
high-order QCD corrections is mandatory to provide accurate theoretical
predictions for the production and annihilation rates, however, this is a
challenging theoretical problem. For most of the channels, the
perturbative corrections are only known to one-loop accuracy~\cite{KMRR}.
The complete ${\cal O}(\alpha_s^2)$ correction is known only for the one-
and two-photon mediated processes
\cite{KPP,CzaMel1,BSS,HoaTeu,PenPiv1,MelYel1,PenPiv2,CzaMel2}.  The
perturbative series, however, shows a slow convergence and the results
suffer from a rather strong dependence on the normalization scale $\nu$
of $\alpha_s(\nu)$. The
resummation of  the large logarithms of the heavy quark velocity to
all orders in $\alpha_s$  has been advocated as a tool to improve the behaviour
of the perturbative expansion for $t\bar t$ threshold production \cite{HMST}.
Currently, the complete next-to-leading logarithmic (NLL) approximation
for the production and annihilation rates is available
\cite{Pin2,HoaSte}.  The first attempt to go beyond the NLL
approximation \cite{HMST} suggested a very good convergence of the
logarithmic expansion.  In particular, an accuracy of 2-3\% was
claimed for the cross section of $t\bar t$ threshold production.
However, subsequent calculations of some
next-to-next-to-leading logarithmic (NNLL)
terms \cite{Hoa}, which had not been taken into account in
Ref.~\cite{HMST},
casted serious doubts on this optimistic estimate.  Thus, the full
calculation of the NNLL corrections, which still remains elusive,
is unavoidable to draw definite conclusions.

In this paper we derive the {\it complete} NNLL result for the
spin dependent part of the heavy quarkonium production annihilation
rates which includes the terms of the form $\alpha_s^{n+2}\ln^n\alpha_s$
for all $n$.  The structure of the paper is as follows.  In the next
section we recall the basic ingredients of the effective theory
description of the production and annihilation rates.  In
Sect.~\ref{run} we derive the nonrelativistic renormalization group
(NRG) equation and in Sect.~\ref{sol} we
present its solution.  The result is applied to $t\bar{t}$ threshold
production, bottomonium and charmonium phenomenology in Sect.~\ref{num}.
Sect.~\ref{sum} contains our conclusions.
Explicit results for the potentials and the analytical solution of the
equations are given in  Appendices A and B, respectively.
Appendix~C includes  the  new results for the high-order logarithmic
corrections to the positronium decay rates.


\section{Effective theory description
  of the production and annihilation rates}
\label{eff}

The perturbative dynamics of the nonrelativistic bound state is
characterised by three well separated scales: the
heavy quark mass $m_q$ (the hard scale), the bound state momentum
$vm_q$ (the soft scale), and the bound state energy $v^2m_q$ (the ultrasoft scale),
where $v \propto\alpha_s \ll 1$ is the velocity of the nonrelativistic heavy quark
for the approximately Coulombic bound state.
A systematic evaluation of the corrections to the bound state parameters
is based on the effective field theory concept of the scale separation
\cite{CasLep}, which is at the heart of the recent progress in the
perturbative QCD bound-state calculations.

The one-photon mediated processes are induced by the electromagnetic
current $j_\mu$. Its space components have the following decomposition in
terms of operators constructed from the nonrelativistic quark and
antiquark two-component Pauli spinors $\psi$ and $\chi$ \cite{BBL}:
\begin{equation}
\bfm{j}=c_v(\nu)\psi^\dagger{\bfm\sigma}\chi+{d_v(\nu)\over6m_q^2}
\psi^\dagger\bfm{\sigma}\mbox{\boldmath$D$}^2\chi
+\ldots,
\label{vcurr}
\end{equation}
where $\nu$ is the renormalization scale, $\bfm{D}$ is the covariant
derivative, ${\bfm\sigma}$ is the Pauli matrix, and the ellipsis stands
for operators of higher mass dimension.  The Wilson coefficients $c_v$
and $d_v$ represent the contributions from the hard modes and may be
evaluated as a series in $\alpha_s$ in
full QCD for free on-shell on-threshold external (anti)quark fields.
We define it through
\begin{eqnarray}
  c_v(\nu) &=& \sum_{i=0}^\infty\left(\alpha_s(\nu)\over
  \pi\right)^i c_v^{(i)}(\nu)\
  \,, \qquad c_v^{(0)}=1\,,
\end{eqnarray}
and similarly for other  coefficients.

The operator responsible for the two-photon $S$-wave  processes in the
nonrelativistic limit is generated by the expansion of the product of
two electromagnetic currents and has the following representation~\cite{BBL}
\begin{equation}
O_{\gamma\gamma}=c_{\gamma\gamma}(\nu)\psi^\dagger\chi
+{d_{\gamma\gamma}(\nu)\over6m_q^2}
\psi^\dagger\mbox{\boldmath$D$}^2\chi
+\ldots,
\label{gcurr}
\end{equation}
which  reduces to the pseudoscalar current in the nonrelativistic limit.
The one-loop corrections to the hard coefficients are known for quite a
long time \cite{KalSar,HarBro}
\begin{eqnarray}
  c_v^{(1)}&=&-2C_F\,,\nonumber\\
  c_{\gamma\gamma}^{(1)}&=&-\left(\frac{5}{2}-\frac{\pi^2}{8}\right)C_F\,,
\end{eqnarray}
where $C_F=(N_c^2-1)/(2N_c)$, $N_c=3$. Let us define the spin ratio for
the production and annihilation of heavy quarkonium ${\cal Q}$ as
\be
{\cal R}_q=
{\sigma(e^+e^-  \rightarrow {\cal Q}(n^3S_1) )\over
\sigma(\gamma\gamma \rightarrow {\cal Q}(n^1S_0))}=
{\Gamma({\cal Q}(n^3S_1)\to e^+e^-)\over
\Gamma({\cal Q}(n^1S_0)\to  \gamma\gamma)}\,.
\ee
The effective theory expression for the spin ratio reads
\be
{\cal R}_q={c_s^2(\nu)\over 3Q_q^2}
{|\psi_n^{v}(0)|^2\over|\psi_n^{p}(0)|^2}+{\cal O}(\alpha_s v^2)\,,
\label{Rdef}
\ee
where $Q_q$ is the quark electric charge,
$c_s(\nu)=c_v(\nu)/c_{\gamma\gamma}(\nu)$, $\psi_n^{(v,p)}(\bfm{r})$ are
the spin triplet (vector) and spin singlet (pseudoscalar) quarkonium
wave functions of the principal quantum number $n$.
The wave functions
describe the dynamics of the nonrelativistic bound state and can be
computed within potential NRQCD (pNRQCD) \cite{PinSot1}.  The latter is
the Schr\"odinger-like effective theory of potential (anti)quarks whose
energy scales like $m_qv^2$ and three-momentum scales like $m_qv$, and
their multipole interaction to the ultrasoft gluons
\cite{KniPen1,BPSV2}.  The contributions of hard and soft modes in
pNRQCD are represented by the perturbative and relativistic correction
to the effective Hamiltonian, which is systematically evaluated order by
order in $\alpha_s$ and $v$ around the leading order (LO) Coulomb
approximation.  The LO Coulomb wave function reads
$\left|\psi^C_n(0)\right|^2=C_F^3\alpha_s^3m_q^3/(8\pi n^3)$. Let us define
\begin{eqnarray}
  \frac{|\psi_n^{v}(0)|^2}{|\psi_n^{p}(0)|^2} &=&
  \rho_n(\nu)\,\,=\,\,\sum_{i=0}^\infty
  \left(\frac{\alpha_s(\nu)}{\pi}\right)^i \rho_n^{(i)}(\nu)
  \,, \qquad \rho_n^{(0)}=1\,.
\end{eqnarray}
The next-to-leading (NLO) contribution $\rho_n^{(1)}$ vanishes since the
corrections to the wave functions are spin-independent at this order.

Starting from ${\cal O}(\alpha_s^2)$, the hard coefficients are infrared
(IR) divergent.  This spurious divergence arises in the process of scale
separation and is canceled against the ultraviolet (UV) one of the
effective-theory result for the wave function at the origin. A powerful
approach to deal with such divergences has been developed in
Refs.~\cite{PinSot2,CMY,Beneke:1999qg,KPSS1}. It is based on
dimensional regularization and the interpretation of the formal expressions
derived from the Feynman rules of the effective theory
in the sense of the threshold expansion \cite{BenSmi,Smi}.
This provides a factorization of the contributions from different
scales.  In the $\overline{\rm MS}$ the
two-loop corrections to $c_v$ are known in analytical form and given
by~\cite{CzaMel1,BSS}
\bea
\left[c_v^{(2)}(\nu)\right]^{\overline{\rm MS}}&=&\left(-{151\over72}
+{89\pi^2\over144}
-{5\pi^2\over6}\ln2-{13\over4}\zeta(3)\right)C_AC_F
+\left({23\over8}-{79\pi^2\over36}\right.
\nn\\
&&
+\left.\pi^2\ln2-{1\over2}\zeta(3)\right)C_F^2
+\left({22\over9}-{2\pi^2\over9}\right)C_FT_F
\nn\\
&&
+{11\over18}C_FT_Fn_l+\left[\beta_0+\pi^2\left({C_A\over 2}
+{C_F \over 3}\right)\right]C_F\ln\left({m_q\over \nu}\right)\,,
\label{cv2}
\end{eqnarray}
where $\beta_0=11C_A/3-4n_lT_F/3$, $C_A=N_c$, $T_F=1/2$, $n_l$ is the
number of light-quark flavours,
$\zeta(3)=1.202057\ldots$ is the value of Riemann's $\zeta$ function,
and $\alpha_s$ is renormalized in the $\overline{\rm MS}$ scheme.
For the two-photon processes the two-loop correction is known in
semi-numerical form \cite{CzaMel2}
\begin{eqnarray}
\left[c_{\gamma\gamma}^{(2)}(\nu)\right]^{\overline{\rm MS}}&=&
-4.79(5)C_AC_F-21.02(10)C_F^2+0.224(1)C_FT_F
+\left(\frac{41}{36}-\frac{13\pi^2}{144}-\frac{2}{3}\ln2 \right.
\nn\\
&&
\left.-\frac{7}{24}\zeta(3)\right)C_FT_Fn_l+\left[\left(\frac{5}{4}
-\frac{\pi^2}{16}\right)\beta_0+\pi^2\left({C_A
\over 2}+{C_F}\right)\right]C_F\ln\left({m_q\over \nu}\right)\,,
\label{cg2}
\end{eqnarray}
where the $gg\to\gamma\gamma$ contribution induced by a light-fermion
box was estimated to be small and was not included in
the result.
Note that the above result depends on the definition of the nonrelativistic
axial current.  The next-to-next-to-leading (NNLO)
correction to the wave functions  ratio in $\overline{\rm MS}$
scheme reads \cite{CzaMel2,KPSS2}
\bea
\left[\rho_n^{(2)}(\nu)\right]^{\overline{\rm MS}}&=&
{2 \over 3}\pi^2C_F^2\left[-\frac{7}{3}+2\Psi_1(n)+2\gamma_E-\frac{2}{n}
+2\ln\left(\frac{C_F\alpha_s(\nu)m_q}{n\nu}\right)\right]\,.
\label{rho2}
\eea
where $\Psi_n(x)=d^n\ln\Gamma(x)/dx^n$,
$\Gamma(z)$ is Euler's $\Gamma$ function,
and $\gamma_E=0.577216\ldots$ is Euler's constant.
However, for the renormalization group analysis it is more convenient to
use the {\it hard matching}  scheme where the nonlogarithmic  part of the
divergent two-loop potential-potential contribution,
which appears in the time-independent perturbation theory for the
nonrelativistic wave function, is shifted to the hard coefficient.
The NNLO corrections in this scheme are obtained
from the  $\overline{\rm MS}$ result by the following shift
\be
\hat\rho_n^{(2)}=  \left[\rho_n^{(2)}\right]^{\overline{\rm MS}}+
\frac{14}{9}\pi^2C_F^2\,, \qquad \hat c^{(2)}_s=
\left[c^{(2)}_s\right]^{\overline{\rm MS}}- \frac{7}{9}\pi^2 C_F^2\,.
\label{hms}
\ee


\section{Renormalization group evolution of $c_s$}
\label{run}

In general, the logarithmic corrections originate from
logarithmic integrals over virtual momenta ranging between the scales
and reveal themselves as the singularities of the effective theory
couplings. The renormalization of these singularities allows one to
derive the equations of the nonrelativistic renormalization group (NRG),
which describe the running of the
effective theory couplings (Wilson coefficients), {\it i.e.} their
dependence on the effective theory cutoffs.  The solution of these
equations sums up the logarithms of the scale ratios.  To derive NRG
equations necessary for the NNLL analysis of the decay rates we rely on
the method developed in Ref.~\cite{Pin2} where, in particular, the
correct NLL result for the decay rates has been obtained for the first
time  (see also Ref.~\cite{HoaSte}).  The NRG equations express the
dependence of the effective theory coupling constants on the IR cutoff.
However, they are derived by studying the UV divergences of the
effective theory
perturbative expressions. In general, one has to consider the soft,
potential and ultrasoft running of the effective theory coupling
constants.  We denote the corresponding cutoffs as $\nu_s$, $\nu_p$ and
$\nu_{us}$, respectively. The soft running is associated with the
divergences of the NRQCD perturbation theory for the potential while the
potential running corresponds to the divergences of the time-independent
perturbation theory for the nonrelativistic Green function in pNRQCD.
As it was first realized in Ref.~\cite{LMR}, $\nu_{us}$ and $\nu_{p}$ are
correlated and the relation between them can be given by
$\nu_{us}=\nu_p^2/m_q$. The matching to the hard contribution
is performed at a generic scale $\nu_h \sim m_q$.

To NNLL approximation the hard coefficient $c_s$ itself has only potential
running.  We compute the corresponding anomalous dimensions by
inspecting the UV singular behaviour of the three-loop effective theory
diagrams computed in Refs.  \cite{KniPen2,KPSS2} (see also
\cite{KniPen3,HilLep,MelYel2}). The NRG equation in the hard matching
scheme is found to be
\be
{d \ln \hat c_s(\nu)  \over d\ln \nu} = {C_F^2\over 3}
\left[2\alpha_{V_s}(\nu)D_{S^2,s}^{(2)}(\nu)
+\alpha_s(\nu){d D_{S^2,s}^{(2)}(\nu) \over d\ln\nu}
\right]\,.
\label{csrge}
\ee
Here $\alpha_{V_s}(\nu)=\alpha_s(\nu)+\ldots$ is the strong
coupling constant defined through the perturbative potential between the
static quark and antiquark in colour-singlet state
\be
V_C(\bfm{k^2})={4\pi C_F\alpha_{V_s}(\bfm{k^2})\over \bfm{k^2}}\,,
\ee
and  $D_{S^2,s}^{(2)}(\nu)=\alpha_s(\nu)+\ldots$ is the Wilson
coefficient of the spin-flip potential in pNRQCD Hamiltonian
\be
V^{(2)}_{S^2}(\nu_p,\nu_s)={4\pi {\bfm S}^2C_FD_{S^2,s}^{(2)}(\nu_p,\nu_s)\over 3 m_q^2}\,,
\ee
with ${\bfm S}=(\bfm{\sigma}_1+\bfm{\sigma}_2)/2$ being the total spin
operator. In Eq.~(\ref{csrge}) we combine the soft and potential running by
setting $\nu_s=\nu_p=\nu$, which is consistent to the order of interest.  Note
that the scheme used in this paper differs from the standard one used for the
evaluation of the Feynman integrals in dimensional regularization in
Ref.~\cite{KPSS2}.

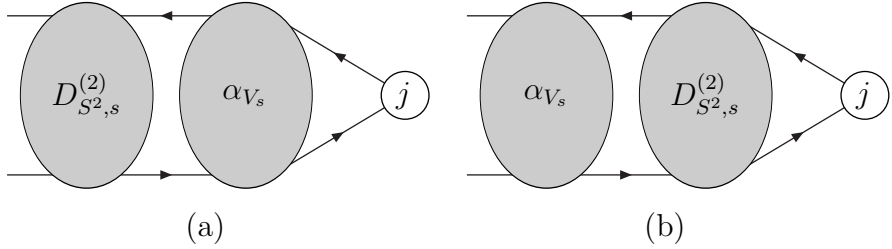
\begin{figure}[t]
\vspace{2em}
\begin{center}
\begin{picture}(170,80)(0,0)
\Line(20,0)(0,0)
\Line(20,60)(0,60)
\ArrowLine(20,0)(100,0)
\ArrowLine(100,60)(20,60)
\ArrowLine(150,30)(100,60)
\ArrowLine(100,0)(150,30)
\GOval(30,30)(35,25)(0){0.8}
\GOval(90,30)(35,25)(0){0.8}
\GCirc(150,30){9}{1}
\Text(150,30)[cc]{${j}$}
\Text(30,30)[cc]{$D_{S^2,s}^{(2)}$}
\Text(75,-20)[cc]{(a)}
\Text(90,30)[cc]{$\alpha_{V_s}$}
\end{picture}
\begin{picture}(170,80)(0,0)
\Line(20,0)(0,0)
\Line(20,60)(0,60)
\ArrowLine(20,0)(100,0)
\ArrowLine(100,60)(20,60)
\ArrowLine(150,30)(100,60)
\ArrowLine(100,0)(150,30)
\GOval(30,30)(35,25)(0){0.8}
\GOval(90,30)(35,25)(0){0.8}
\GCirc(150,30){9}{1}
\Text(150,30)[cc]{${j}$}
\Text(90,30)[cc]{$D_{S^2,s}^{(2)}$}
\Text(75,-20)[cc]{(b)}
\Text(30,30)[cc]{$\alpha_{V_s}$}
\end{picture}
\vspace{10mm}
\caption{\small The pNRQCD diagrams contributing to the
NNLL running of $c_s$. The shaded oval represents the soft running of the spin-flip
or Coulomb potential and $j$ is nonrelativistic vector or pseudoscalar
current.}
\label{fig1}
\end{center}
\end{figure}

Let us now discuss the structure of the NRG
equation~(\ref{csrge}) in more detail.  The first term on the right-hand side
originates from the diagram (a) in Fig.~\ref{fig1}. The leading logarithmic
(LL) soft running of $\alpha_{V_s}$ and $D_{S^2,s}^{(2)}$ is responsible for
the NLL evolution of $c_s$.  To get the NNLL result for the hard coefficient
we need the NLL running of this quantities.  The one of $\alpha_{V_s}$ is well
known and can be found {\it e.g.} in Ref.~\cite{Pin2}.  For $D_{S^2,s}^{(2)}$
it has been recently obtained in Refs.~\cite{KPPSS,PPSS}.  The analytical
result for the NLL evolution of the spin-flip potential is presented in
Ref.~\cite{PPSS} where, however, some details of the analysis have been
skipped.  This gap is filled in Appendix~A, where explicit results for the
contributing potentials are presented.

The second term on the right-hand side of Eq.~(\ref{csrge})
originates from the diagram (b) of
Fig.~\ref{fig1}. It starts to contribute to $c_s$ at NNLL. Thus the
LL expression for $D_{S^2,s}^{(2)}(\nu)$ can be used, which is given
by~\cite{Pin2,MS}
\be
{d\over d\ln\nu}D_{S^2,s}^{(2)}(\nu)={\alpha^2_s(\nu)\over\pi}
\gamma_s^{(1)} c_F^2(\nu)\,, \qquad  \gamma_s^{(1)}=-{\beta_0 \over 2}+{7
  \over 4}C_A\,.
\label{Dsll}
\ee
$c_F$ is the effective Fermi coupling, which is needed with LL accuracy, and
reads $c_F(\nu_h)=z^{-C_A}$, where
$z=\left(\alpha_s(\nu)/\alpha_s(\nu_h)\right)^{1/\beta_0}$.

The initial condition for the NNLL solution of the NRG evolution is fixed by
the two-loop value of $c_s(\nu_h)$ at a hard matching scale $\nu_h\sim
m_q$. The subsequent evolution of $c_s(\nu)$ down to $\nu\sim
\alpha_sm_q$ resums the logarithms of the coupling constant. The form
of the NRG equation and the matching conditions is scheme dependent.  A
change of the scheme $\hat c_s^{(2)}(\nu_h)\to c_s^{(2)}(\nu_h)=
\hat c_s^{(2)}(\nu_h)-\delta c_s\, \pi^2C_F^2$ requires an additional full
derivative term to be added to the right hand side of Eq.~(\ref{csrge})
\be
\delta c_s\, C_F^2 {d  \over d\ln\nu}
\left(\alpha_{V_s}(\nu)D_{S^2,s}^{(2)}(\nu)\right)=\delta c_s \, C_F^2
{\alpha^2_s(\nu)\over\pi}\left(-{\beta_0\over 2}D_{S^2,s}^{(2)}(\nu)
+ \gamma_s^{(1)}\alpha_s(\nu) c_F^2(\nu)\right)
+\ldots\,,
\ee
where $\delta c_s$ specifies the scheme and the ellipses indicated higher orders
in $\alpha_s$. In particular we have $[\delta c_s]^{\overline{\rm MS}}=-7/9$.
Note that the coupling $c_F$
appears in Eq.~(\ref{csrge}) in combination with the factor $\gamma_s^{(1)}$.

To get the NNLL approximation for the production and annihilation rates
we formally have to take into account also the LL running of $d_{v}$ and
$d_{\gamma\gamma}$. These hard coefficients, however, have identical LL
ultrasoft running which cancels out in the spin ratio.


\section{Solution of the renormalization group equation}
\label{sol}

The solution of the renormalization group equation~(\ref{csrge})
is of the following form
\be
\hat c_s(\nu)=\hat c_s(\nu_h)
  e^{\alpha_s(\nu_h)\Gamma_{\hat c_s}^{\rm NLL}(\nu)
    +\alpha_s^2(\nu_h)\Gamma_{\hat c_s}^{\rm NNLL}(\nu)+\cdots}
  \,.
\label{csnnl}
\ee
The expression of $\Gamma_{\hat c_s}^{\rm NLL}$ is known \cite{Pin2,HoaSte}
\bea
\Gamma_{\hat c_s}^{\rm NLL}(\nu)&=&
{2\pi C_F^2 (2\beta_0 -7C_A)
     \over 3 (\beta_0-2 C_A)^2 }\left(1-z^{\beta_0-2 C_A}\right)
-{2\pi C_F^2 C_A\over \beta_0 (\beta_0-2C_A)}\ln(z^{\beta_0})\,.
\label{Gammanll}
\eea
The result for the NNLL function can also be obtained
analytically\footnote{For the practical calculation we use
  $\nu_{us}=\nu^2/\nu_h$, which is sufficient for the
  accuracy of our calculation.} and
can be cast in the form
\be
\Gamma_{\hat c_s}^{\rm NNLL}(\nu)= \pi^2 \sum_{i=1}^{19}
A_i f_i(z)
\,,
\label{Gammannll}
\ee
where the coefficients $A_i$ and functions $f_i(z)$ are given in
Appendix~B.  To get the NNLL approximation for the spin ratio one has to
take into account the NNLL contribution to the wave function, which is
given by
\be
\rho_n^{\rm NNLL}(\nu)=1+\frac{1}{\pi^2}
\left(\alpha_{V_s}(\nu)D_{S^2,s}^{(2)}(\nu)\right)^{\rm LL}
\hat\rho_n^{(2)}(\nu)
\,.
\label{rhonnll}
\ee
It cancels the NLL $\nu$-dependence of $c_s$.  Eqs.~(\ref{csnnl})
and~(\ref{rhonnll}) give the NNLL approximation to the spin ratio. By
expanding the resummed expression we reproduce the known results for the
${\cal O}(\alpha_s^2)$ \cite{CzaMel2}, ${\cal  O}(\alpha_s^3\ln^2\alpha_s)$
\cite{KniPen2}, and ${\cal  O}(\alpha_s^3\ln\alpha_s)$ \cite{KPSS2}
terms.  After including the one-photon annihilation contribution, the
Abelian part of our result reproduces the
spin-dependent corrections of order $\alpha^3\ln\alpha$
to the positronium decay rates~\cite{KniPen3,MelYel2}.
The use of the NRG allows us to derive the
higher order logarithmic corrections for positronium. The explicit
results for the ${\cal O}(\alpha^4\ln^2\alpha)$ and ${\cal
  O}(\alpha^5\ln^3\alpha)$ terms are given in Appendix~C.


\section{Heavy quarkonium phenomenology}
\label{num}

\begin{figure}[t]
\begin{center}
\epsfxsize=\textwidth
\epsffile{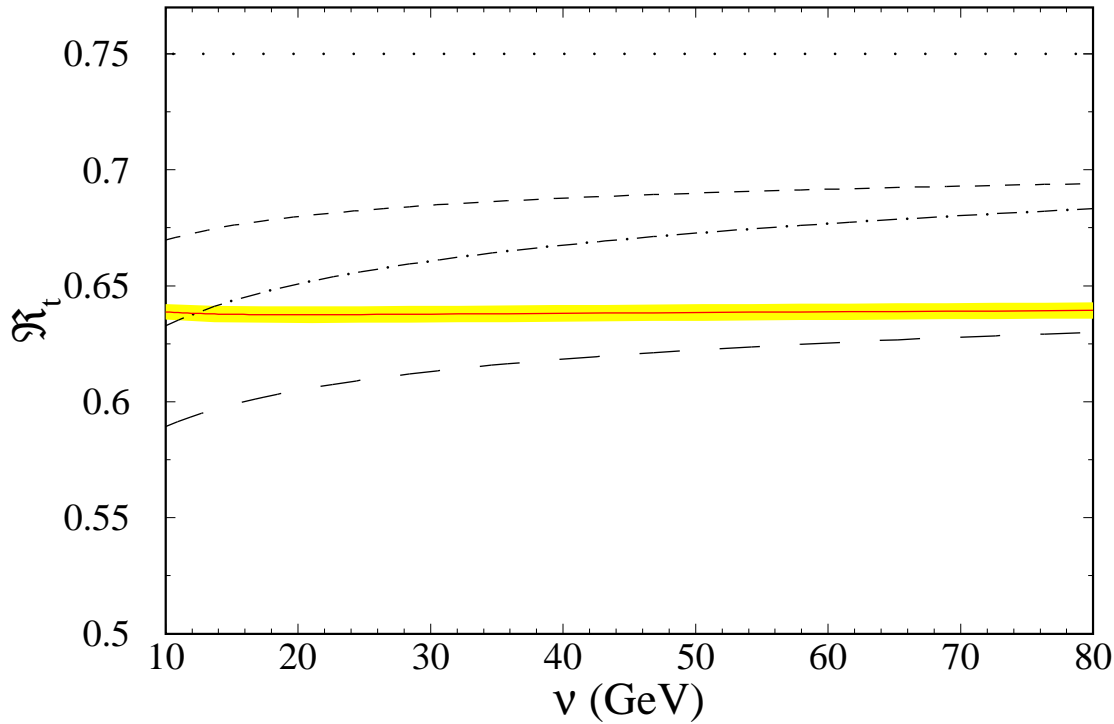}
\end{center}
\caption{\label{figt} The spin  ratio  as the function of the
renormalization scale $\nu$ in LO$\equiv$LL (dotted line), NLO (short-dashed
line), NNLO (long-dashed line), NLL (dot-dashed
line), and NNLL (solid line) approximation for the (would be)
toponium ground state with $\nu_h=m_t$. For the NNLL result the band
reflects the errors due to $\alpha_s(M_Z)=0.118\pm 0.003$. Note that
for the vertical axis the zero is suppressed.}
\end{figure}

\begin{figure}[t]
\begin{center}
\epsfxsize=\textwidth
\epsffile{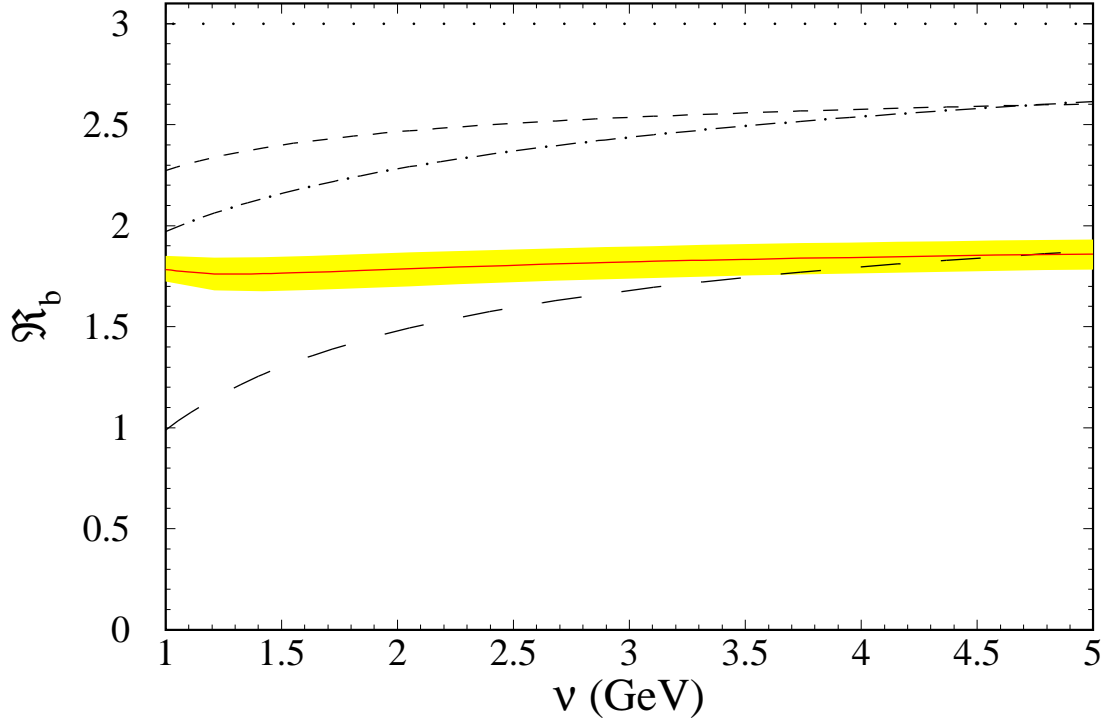}
\end{center}
\caption{\label{figb} The spin  ratio  as the function of the
renormalization scale $\nu$ in LO$\equiv$LL (dotted line), NLO (short-dashed
line), NNLO (long-dashed line), NLL (dot-dashed
line), and NNLL (solid line) approximation for the bottomonium
ground state with $\nu_h=m_b$. For the NNLL result the band reflects
the errors due to $\alpha_s(M_Z)=0.118\pm 0.003$.}
\end{figure}

\begin{figure}[t]
\begin{center}
\epsfxsize=\textwidth
\epsffile{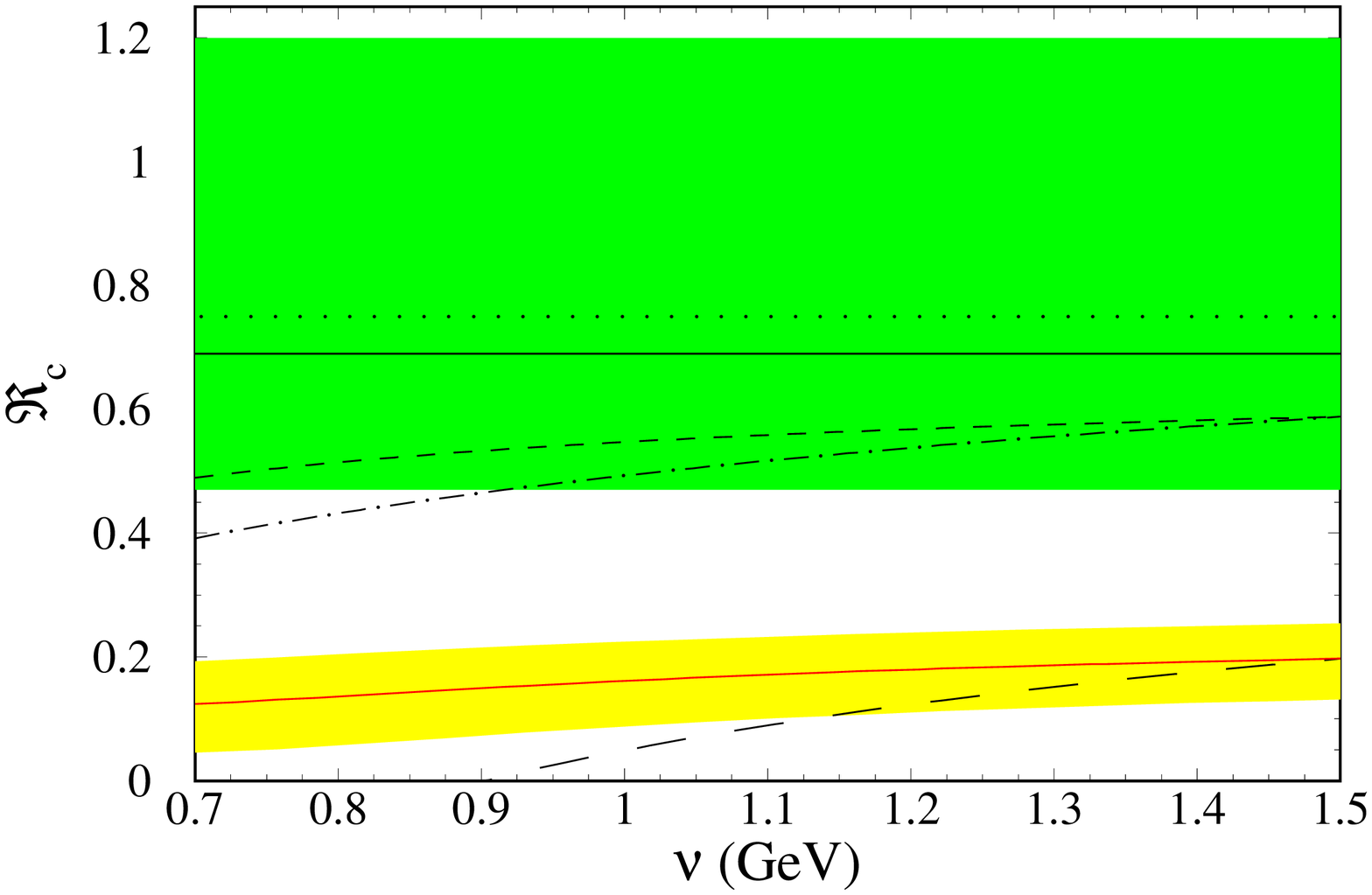}
\end{center}
\caption{\label{figc}  The spin  ratio  as the function of the
renormalization scale $\nu$ in LO$\equiv$LL (dotted line), NLO (short-dashed
line), NNLO (long-dashed line), NLL (dot-dashed
line), and NNLL (solid line) approximation for the charmonium
ground state with $\nu_h=m_c$.  For the NNLL result the lower (yellow)
band reflects
the errors due to $\alpha_s(M_Z)=0.118\pm 0.003$. The upper (green) band
represents the experimental error of the ratio \cite{Hag} where the central
value is given by the horizontal solid line.}
\end{figure}

For the numerical estimates, we take $m_c=1.5$ GeV,
$m_b=M\left(\Upsilon\right)/2$ and $m_t=175$~GeV, which is sufficient at
the order of interest. Furthermore, we take $\alpha_s(M_Z)$ as an input
and run with four-loop accuracy down to the matching scale $\nu_h$ to
ensure the best precision.  Below the matching scale, the running of
$\alpha_s$ is used according to the logarithmic precision of the
calculation in order not to include the corrections beyond the NNLL
accuracy. By the same reason we expand the decay rates ratio in
$\alpha_s(\nu_h)$ up to the NNLL accuracy.
In Figs.~\ref{figt}, \ref{figb} and \ref{figc}, the spin ratio
is plotted as a function of $\nu$ in the various logarithmic and
fixed-order approximations for the (hypothetical) toponium, bottomonium and
charmonium ground states, respectively.  As we see, in the second order the
convergence and stability of the result with respect to the scale
variation is substantially improved if one switches from the
fixed-order to the logarithmic expansion. We want to remark that the
$\nu$ dependence of the NLL approximation is slightly worse than at
NLO. This is due to the artificially small $\nu$ dependence at NLO
which is likely due to the fact that at this order only the hard scale
enters.

Let us first consider the top quark case.  The relatively large top
quark width smears out the Coulomb-like poles below the threshold
leaving a single well-pronounced resonance with the properties mainly
determined by the ``would be'' toponium ground state parameters.  Thus
our result can be applied to the ratio of the cross sections of the
resonance $e^+e^-\to t\bar t$ and $\gamma\gamma\to t\bar t$ production.
As one can see in Fig.~\ref{figt}
the logarithmic expansion shows perfect convergence and the NNLL
correction vanishes at the scale $\nu\approx 13$~GeV, which is close to
the physically motivated scale of the inverse Bohr radius
$\alpha_sm_t/2$. For illustration, at the scale of minimal sensitivity,
$\nu=20.2$ GeV, we have
\be
{\sigma_{\rm res}(e^+e^-\rightarrow  t\bar t)\over\sigma_{\rm res}
(\gamma\gamma\to t\bar t)}={1\over 3Q_t^2}\left(1-0.132-0.018\right)
\,.
\ee
Note that the perturbative expansion for the ground state energy, which
is known to ${\cal O}(\alpha_s^3)$ \cite{PenSte}, shows as similar nice
property.  However, it is not clear if the nice behaviour of the
logarithmic expansion also holds for the spin-independent part of the
threshold cross section. A possible problem is connected to
the ultrasoft contribution,
which is enhanced by the larger value of $\alpha_s$ at the ultrasoft
scale. Whereas it is suppressed in the spin ratio by the fifth power of $\alpha_s$,
for the spin-independent part it already contributes at
${\cal O}(\alpha_s^3)$ and can destabilize the expansion.

For bottomonium, the logarithmic expansion shows nice convergence and
stability (c.f. Fig.~\ref{figb})
despite the presence of ultrasoft contributions with
$\alpha_s$ normalized at a rather low scale $\nu^2/m_b$.
At the same time, the perturbative corrections are important
and reduce the leading order result by approximately $41\%$.
For illustration, at the
scale of minimal sensitivity, $\nu=1.295$~GeV, we have the following series:
\be
{\Gamma(\Upsilon(1S) \rightarrow
  e^+e^-)\over\Gamma(\eta_b(1S)\rightarrow\gamma\gamma)}
={1\over 3Q_b^2}\left(1-0.302-0.115\right)\,.
\ee
In contrast, the fixed-order expansion blows up at the scale
of the inverse Bohr radius.

So far we have discussed the perturbative corrections to the Coulomb-like
quarkonium.  However, in contrast to the $t\bar{t}$ system,
for bottomonium nonperturbative contributions can be
important. In our case the interaction of the quark-antiquark pair to the
nonperturbative gluonic field is suppressed by $v$ through
the multipole expansion.  In every order of the multipole expansion the
nonperturbative contribution is given by the convolution of the quantum
mechanical Green function with a nonlocal nonperturbative gluonic
correlator. In general, we know little about the structure of these
quantities.  If $v^2m_q\gg \Lambda_{QCD}$, it can be investigated by the
method of vacuum condensate expansion \cite{Vol,Leu}. The resulting
series, however, is not expected to converge well in our case and
suffers from large numerical uncertainties \cite{TitYnd2,Pin1}.
Nevertheless, it is possible to estimate the nonrelativistic suppression
of the leading nonperturbative effect. For the ratio of the decay rates the
nonperturbative contribution emerges via the ${\cal O}(v^4)$
chromomagnetic dipole interaction and the interference between the
${\cal O}(v^2)$ chromoelectric dipole interaction and the leading ${\cal
  O}(v^2)$ spin-flip Fermi potential.  Hence it is suppressed by the
factor $v^4$.  Within the power counting assumed in this paper
it only contributes in the N$^4$LL approximation, far beyond the precision
of our computation. Note that the nonperturbative contribution to the
decay rates ratio
is suppressed by a factor $v^2$ in comparison to the
binding energy and decay rates, where the leading nonperturbative effect
is due to chromoelectric dipole interaction.  Thus the renormalization
group improved result allows for very accurate theoretical evaluation of
the spin ratio and, by using the available experimental data on the
$\Upsilon$ meson as input, we can predict the production and
annihilation rates of the yet undiscovered $\eta_b$ meson with high
precision.  In particular, we predict the $\eta_b(1S)$ decay rate using
the experimental value for the $\Upsilon(1S)$ decay rate
\be
\label{etabgg}
\Gamma(\eta_b(1S) \rightarrow \gamma\gamma)=0.659\pm 0.089 ({\rm th.})
{}^{+0.019}_{-0.018} (\delta \alpha_{\rm s})\pm 0.015 ({\rm exp.})\; {\rm keV}
\,,
\ee
where we have taken $\nu=1.295$ GeV, the scale of minimal sensitivity,
for the central value, the difference between the NLL and NNLL result
for the theoretical error and $\alpha_{\rm s}(M_Z)=0.118 \pm 0.003$. The
last error in Eq.~(\ref{etabgg}) reflects the experimental error of
$\Gamma(\Upsilon(1S) \rightarrow e^+e^-)=1.314\pm 0.029$ keV \cite{Hag}.
This value considerably exceeds the result for the absolute value of the
decay width obtained in Ref.~\cite{Pin3} on the basis of the full NLL
analysis including the spin-independent part.
This can be a signal of
slow convergence of the logarithmic expansion for the spin-independent
contribution which is more sensitive to the dynamics of the bound state
and in particular to the ultrasoft contribution as it has been
discussed above. On the other hand, the renormalon effects  \cite{BC}
could produce some systematic errors in the pure perturbative
evaluations of the production/annihilation rates. The problem is
expected to be more severe for the charmonium case discussed below.

We would also like to remark that the one-loop result for $\nu=m_b$
overshoots the NNLL one by approximately $30\%$.  This casts some doubts
on the accuracy of the existing $\alpha_s$ determination from the
$\Gamma(\Upsilon\to{\rm light~hadrons})/\Gamma(\Upsilon\to e^+e^-)$
decay rates ratio, which gives $\alpha_s(m_b)=0.177\pm 0.01$, well below
the ``world average'' value \cite{Hag}.  The theoretical uncertainty in
the analysis is estimated through the scale dependence of the one-loop
result.  Our analysis of the photon mediated annihilation rates
indicates that the actual magnitude of the higher order corrections is
most likely quite beyond such an estimate and the theoretical
uncertainty given in \cite{Hag} should be increased by a factor of
two. This brings the result for $\alpha_s$ into $1\sigma$ distance from
the ``world average'' value.

For the charmonium, the NNLO approximation becomes negative at an
intermediate scale between $\alpha_sm_c$ and $m_c$ (c.f. Fig.~\ref{figc})
and the use of the
NRG is mandatory to get a sensible perturbative approximation. The NNLL
approximation has good stability against the scale variation but the
logarithmic expansion does not converge well.  This is the main factor
that limits the theoretical accuracy since the nonperturbative
contribution is expected to be under control. For illustration, at the
scale of minimal sensitivity, $\nu=0.645$ GeV, we obtain
\be
{\Gamma(J/\Psi(1S)\rightarrow
  e^+e^-)\over\Gamma(\eta_c(1S)\rightarrow\gamma\gamma)}
={1\over 3Q_c^2}\left(1-0.513-0.340\right)\,.
\ee
The central value of our NNLL result is $2\sigma$ below the experimental
value. The discrepancy may be explained by the large higher order
contributions. This should not be surprising because of the rather large value
of $\alpha_s$ at the inverse Bohr radius of charmonium.  For the charmonium
hyperfine splitting, however, the logarithmic expansion converges well and the
prediction of the NRG is in perfect agreement with the experimental data.
Thus one can try to improve the convergence of the series for the
production/annihilation rates by accurately taking into account the
renormalon-related contributions. One point to note is that with a potential
model evaluation of the wave function correction, the sign of the NNLO term is
reversed in the charmonium case \cite{CzaMel2}. At the same time the
subtraction of the pole mass renormalon from the perturbative static potential
makes explicit that the potential is steeper and closer to lattice and
phenomenological potential models \cite{sum}. Therefore, the incorporation of
higher order effects from the static potential may improve the agreement with
experiment. In any case, if we estimate the theoretical uncertainty as the
difference of the NNLL and the NLL result at the soft scale $\alpha_sm_c$, the
theoretical and experimental values agree within the error bars.


\section{Summary}
\label{sum}

To conclude, we have resummed to all orders in the perturbative expansion the
next-to-next-to-leading logarithms in the ratio of the one-photon mediated
production and annihilation rate of the spin-triplet heavy quarkonium state to
the two-photon one of the spin-singlet state.  This constitutes the first
complete NNLL result for the production and annihilation of heavy quarkonia.
The use of the NRG improves the convergence of the perturbative series and
stabilizes the result with respect to the scale variation. This allows us to
obtain extremely accurate predictions for the $t\bar t$ threshold production.
However, we still cannot draw a definite conclusion about the accuracy of the
NNLL approximation to the spin independent part of the threshold production
cross section, which could be worse due to the ultrasoft contribution.

For the bottomonium case we found convergence of the logarithmic expansion at
the physically motivated scale of inverse Bohr radius and very weak scale
dependence of the NNLL approximation. Using these results we can give
predictions for the yet to be discovered $\eta_b$ meson two-photon
production/annihilation rates.  The magnitude of the higher order corrections
questions the reported accuracy of $\alpha_s$ determination from the
$\Gamma(\Upsilon\to{\rm light~hadrons})/ \Gamma(\Upsilon\to e^+e^-)$ ratio
based on the one-loop theoretical analysis.

For the charmonium annihilation the NRG is mandatory to bring physical sense
to the perturbative result because the NNLO approximation becomes negative at
an intermediate scale.  Though the NNLL approximation has very good stability
with respect to the scale variation, the convergence of the logarithmic
expansion is slow and essentially limits the accuracy of the perturbative
prediction. Taking into account the large uncertainties of our result and the
experimental data there is agreement and a lower value for the ratio seems to
be favoured by our result.

As a by-product of the heavy quarkonium analysis we have obtained novel
results for positronium: the complete $O(\alpha^4\ln^2\alpha)$ and
$O(\alpha^5\ln^3\alpha)$ corrections to the ratio of the ortho- and
parapositronium decay rates.


\vspace{5mm}
\noindent
{\bf Acknowledgments:}\\
The work of A.A.P. was supported in part by BMBF Grant No.\ 05HT4VKA/3 and SFB
Grant No. TR 9. The work of A.P. was supported in part by MCyT and Feder
(Spain), FPA2001-3598, by CIRIT (Catalonia), 2001SGR-00065 and by the EU
network EURIDICE, HPRN-CT2002-00311. The work of V.A.S. was supported in part
by RFBR Project No. 03-02-17177, Volkswagen Foundation Contract No. I/77788,
and DFG Mercator Visiting Professorship No. Ha 202/1. M.S. was supported by
HGF Grant No. VH-NH-008.


\section*{Appendix A:
Nonrelativistic renormalization group
evolution of the spin-flip potential}

In order to obtain $c_s$ to NNLL accuracy one needs the matching coefficient
of the spin-flip potential, $D_{S^2,s}^{(2)}$, to NLL order, which has been
considered in Ref.~\cite{PPSS}. In this Appendix we provide additional
details.

The soft running of  $D_{S^2,s}^{(2)}$ to NLL approximation
is discussed in detail in Ref.~\cite{PPSS}.
The NRG evolution of the spin-flip potential
also includes the potential and ultrasoft running at NLL, which is described below.
To compute the potential running, we inspect all operators that induce
spin-dependent UV
divergences in the time-independent perturbation theory contribution with one
and two potential loops.  They include
\begin{itemize}
\item[(i)] the tree-level ${\cal O}(v^4)$ operators,
\item[(ii)] the one-loop ${\cal O}(\alpha_sv^3)$ operators, and
\item[(iii)] the ${\cal O}(v^2,\alpha_sv)$ operators,
\end{itemize}
which we in turn discuss in the following.
In Appendix~A.4 we discuss the three-loop ultrasoft-potential contribution and,
finally, in Appendix~A.5 we provide the NRG equation for the potential running.


\subsection*{A.1 Tree-level potentials of order $v^4$}

The operators   relevant  for our calculation read
\bea
V_{S^2,1}^{(4)}&=&\pi  C_F\alpha_s {c_S^{(1)}c_S^{(2)}\over 4m_1^2m_2^2}
{1 \over{\bfm k}^2}\bfm{\sigma}_1\cdot({\bf k}\times{\bfm p})\bfm{\sigma}_2
\cdot({\bfm k}\times{\bfm p})\,,
\label{Vt1}\\
V_{S^2,2}^{(4)}&=&-\pi C_F\alpha_s{c_F^{(1)}c_F^{(2)}\over
4m_1^2m_2^2}{({\bfm p}^2-{\bfm p'}^2)^2 \over {\bfm k}^2}
\left(\bfm{\sigma}_1\cdot \bfm{\sigma}_2
-{\bfm{\sigma}_1\cdot{\bfm k}\bfm{\sigma}_2\cdot{\bfm k}\over{\bfm k}^2}\right)\,,
\label{Vt2}\\
V_{S^2,3}^{(4)}&=&\pi  C_F\alpha_s {{\bfm p}^2-{\bfm p'}^2 \over 2{\bfm k}^2}
\left[{c_S^{(1)}c_F^{(2)} \over 4m_1^3m_2}
(\bfm{\sigma}_1\times ({\bfm p}+{\bfm p}'))\cdot(\bfm{\sigma}_2\times
{\bfm k})+(1\leftrightarrow 2)\right]\,,
\label{Vt3}\\
V_{S^2,4}^{(4)}&=&-{\pi  C_F\alpha_s  \over 8{\bfm k}^2}
\left[{c_{pp'}^{(1)}c_F^{(2)} \over m_1^3m_2}\bfm{\sigma}_1\cdot ({\bfm p}+{\bfm p}')
\left(\bfm{\sigma}_2\cdot({\bfm p}+{\bfm p}') {\bfm k}^2+({\bfm p}^2-{\bfm p'}^2)
\bfm{\sigma}_2\cdot {\bfm k}\right)\right.
\nn\\
&&
+(1\leftrightarrow 2)\Bigg]\,.
\label{Vt4}
\eea
Here $\bfm p$ and $\bfm{p}'=\bfm{p}+\bfm{k}$ are the momentum of
incoming and outgoing quark,  we consider the general case with
quark and antiquark of different masses $m_1$ and $m_2$, and adopt the
standard notations for the NRQCD coupling constants $c_F$, {\it etc.}
(see, {\it e.g.}  Ref.~\cite{Man}).  The operators in
Eqs.~(\ref{Vt1})-(\ref{Vt3}) can be inferred from the analysis of the
hyperfine splitting QED \cite{CMY} with a trivial adjustment of the
colour structure.  Eq.~(\ref{Vt1}) corresponds to Eq.~(13) of
Ref.~\cite{CMY}.  The potential~(\ref{Vt2}) results from the expansion
of the transverse gluon propagator in the energy transfer up to $k_0^2$
with subsequent use of the Coulomb equation of motion \cite{PinSot2}
\be
k_0^2 = -{({\bfm p}^2-{\bfm p'}^2)^2 \over 4m_1m_2}\,.
\ee
It reproduces the retardation effect given by Eq.~(32) of
Ref.~\cite{CMY}.

The potentials proportional to $c_Sc_F$ appear both at
tree and one-loop level.  There
are two $c_S$ NRQCD vertices that contribute to our calculation. One of
them is proportional to $A_0{\bf A}$.  It is responsible half of
the contribution proportional to $c_Sc_F$ in Eq.~(\ref{V1l}) of the
next subsection.  The other involves the time derivative $\partial_0{\bf A}$.
To compute this contribution we perform a field redefinition in the
NRQCD Lagrangian which in the lowest order is equivalent to using the
equations of motion. In this way we obtain two new vertices.  The first
vertex is proportional to $({\bf p}^2/m){\bf A}$ and produces the
potential in Eq.~(\ref{Vt3}), which agrees with the first term in
Eq.~(22) of Ref.~\cite{CMY}.  The second vertex is proportional to
$A_0{\bf A}$ and is responsible for the second half of the
$c_Sc_F$ contribution of the one-loop operator in Eq.~(\ref{V1l}).

The potential~(\ref{Vt4}) was not considered in Ref.~\cite{CMY} since in QED
$c_{p'p}={\cal O}(\alpha)$ and thus $V_{S^2,4}^{(4)}$ gives at most
corrections of order $\alpha^3$ to the hyperfine splitting.  Moreover,
in QED this potential contributes only to the NNLL running of the
spin-flip operator as the coupling $c_{p'p}$ has no anomalous dimension.
However, in QCD $c_{p'p}$ does run and we have to take it into account.

There also exist ${\cal O}(v^4)$ potentials including the product of
$c_{W_2}c_F$ or $c_{W_1}c_F$. They are, however, proportional to $({\bfm
  p}\cdot {\bfm p'})(\bfm{\sigma}_1 \cdot \bfm{\sigma}_2)$ and $({\bfm
  p}^2+{\bfm p'}^2)(\bfm{\sigma}_1 \cdot \bfm{\sigma}_2)$, respectively,
which do not produce potential divergences to the order of interest.

To get the NLL approximation for the spin-flip potential we need the LL
running of the ${\cal O}(v^4)$ potentials, which is determined by the
soft running of the NRQCD coupling constants.  They are either known
\cite{Pin2} or can be deduced using reparameterization invariance
\cite{Man}.


\subsection*{A.2 One-loop potentials of order  $\alpha_sv^3$}

The operators   relevant  for our calculation are
\begin{eqnarray}
V^{(3)}_{S^2}&=&
-{1 \over 24}(\pi\alpha_s)^2 C_F(4 C_F-C_A)\left({c_F^{(1)}\over m_1}
{c_S^{(2)}\over m_2^2}+{c_S^{(1)}\over m_1^2}{c_F^{(2)}\over m_2}
\right) |{\bfm k}|\bfm{\sigma}_1\cdot \bfm{\sigma}_2
\nn\\
&&
+{1 \over 24}(\pi\alpha_s)^2 C_F C_A
{c_F^{(1)}\over m_1}{c_F^{(2)}\over m_2} {1 \over |{\bfm k}|}
\left({{\bfm p}^2+{\bfm p'}^2 \over 2m_r}-2E-{{\bfm k}^2 \over 2m_r}
\right)\bfm{\sigma}_1\cdot \bfm{\sigma}_2
\nn\\
&&
-{1 \over 96}(\pi\alpha_s)^2 C_FC_A\left(
{c_F^{(1)}\over m_1}{c_F^{(2)2}\over m_2^2}
+{c_F^{(1)2}\over m_1^2}{c_F^{(2)}\over m_2}
\right) |{\bfm k}|\bfm{\sigma}_1\cdot \bfm{\sigma}_2\,,
\label{V1l}
\end{eqnarray}
where $E$ is the two-particle energy.
The Abelian piece reproduces  Eq.~(34) of Ref.~\cite{CMY} whereas the
non-Abelian part is new.
Note that the second line gives no NLL contribution  as
the term proportional to $E$ develops no logarithm and the remaining two terms
cancel each other. We need the NLL running of this potential, which is
inherited through the LL soft running of the NRQCD coupling constants.


\subsection*{A.3  Double insertion of ${\cal O}(v^2,\alpha_sv)$ operators}

The basis of the ${\cal O}(v^2,\alpha_sv)$  operators is well known from
the  lower order calculations  and can, {\it e.g.} be found in
Refs. \cite{BPSV1,KPSS1}.
To get the NLL contribution to the spin-flip potential from the
double insertion of these operators  we need the LL soft and ultrasoft
running of the ${\cal O}(v^2)$ terms and NLL soft and ultrasoft running of
${\cal O}(\alpha_sv)$ operators, which are  known for the equal mass case
\cite{Pin2}.
For the  non-equal mass case, most of the results can be trivially generalized
(c.f. Ref.~\cite{Pin2}) except for $D_{d,s}^{(2)}$, which
depends on the four-fermion  couplings of NRQCD. Among these couplings
only for $d_{vs}$ a nontrivial change is necessary. The corresponding NRG
equation can be found in Ref.~\cite{Pin2} and its solution reads
\begin{eqnarray}
  d_{vs}(\nu)&=&
  d_{vs}(\nu_h)
  -\alpha_s(\nu_h)
  \left[4 C_F-3{C_A \over 2}
    -{5 \over 4}C_A\left(\frac{m_1}{m_2}+\frac{m_2}{m_1}\right)
  \right]{2\pi\over\beta_0}
  \left(z^{\beta_0}-1 \right)
  \nn\\
  &&
  -\alpha_s(\nu_h)
  \left(\frac{m_1}{m_2}+\frac{m_2}{m_1}\right)
  {27C_A^2 \over 9C_A+8T_Fn_l}{\pi \over 2\beta_0}
  \left\{-\frac{5 C_A + 4 T_F n_l}{4 C_A + 4T_F n_l}\right.
  \nn\\
  &&
  \times {\beta_0 \over \beta_0-2C_A}\left(z^{\beta_0-2 C_A}-1\right)
  +\frac{C_A +16  C_F - 8 T_F n_l}{2(C_A-2T_F n_l)}\left(z^{\beta_0}-1\right)
  \nn\\
  &&
  +\frac{-7 C_A^2 + 32 C_A  C_F - 4 C_A T_F n_l +32  C_F T_F
    n_l}{4(C_A + T_F n_l)(2 T_F n_l-C_A)}
  \nn\\
  &&
  \times
  {3\beta_0 \over 3\beta_0+4T_Fn_l-2C_A}\left( z^{\beta_0+4 T_F n_l/3 - 2C_A/3}-1\right)
  \nn\\
  &&
  +{8T_Fn_l \over 9C_A}\left[{\beta_0 \over \beta_0-2C_A}\left(z^{\beta_0-2C_A}-1\right)
    +\left({20 \over 13}+{32 \over 13}{ C_F \over C_A}\right)\right.
  \nn\\
  &&
  \left.\left.\times
      \left(\left[z^{\beta_0}-1\right]-{6\beta_0 \over 6\beta_0-13C_A}
        \left[z^{\beta_0-{13C_A \over 6}}-1\right]\right)\right]\right\}\,.
\end{eqnarray}


\subsection*{A.4 Three-loop ultrasoft-potential running}

Another NLL contribution  to the spin-flip potential comes from the
three-loop NLL ultrasoft-potential running of  a  ${\cal O}(v^2)$
spin-flip operator. The contribution is
related to the   chromomagnetic dipole gluon exchange which generate the
potential of the following form in the coordinate representation
\be
V^{(2)}_{S^2,1/r^3}={4 C_F{\bf S}^2 \over 3m_1m_2}V_o(V_o-V_s)^2
D^{(2)}_{S^2,1/r^3}\,,
\label{vus}
\ee
where
\bea
V_s &=& - C_F\frac{\alpha_s}{|\bfm {r}|}\,,
\nn\\
V_o &=& \left(\frac{C_A}{2}- C_F\right)\frac{\alpha_s}{|\bfm {r}|}\,.
\eea
The ultrasoft  running of the potential is determined by the NRG equation
\be
{{\rm d}V^{(2)}_{S^2,1/r^3}(\nu_{us})\over {\rm d}\ln\nu_{us}}
=c_F^2(\nu_{us}){\alpha_s(\nu_{us}) \over \pi}{2 C_F \over
3m_1m_2}V_o(V_o-V_s)^2\,,
\ee
and the solution for the Wilson coefficient  reads
\be
D^{(2)}_{S^2,1/r^3}(\nu_{us})=
\frac{1}{2C_A}\left[\left(\frac{\alpha_s(\nu_h)}{\alpha_s(\nu_{us})}\right)^{
2C_A/\beta_0}
-\left(\frac{\alpha_s(\nu_h)}{\alpha_s(1/|\bfm{r}|)}\right)^{2C_A/\beta_0}\right
]\,.
\ee
The potential~(\ref{vus}) is singular at $|\bfm {r}|\to 0$
and should be understood as the Fourier transform
of $\ln(|\bfm {k}|/\nu_p)$.


\subsection*{A.5 NRG equation for potential running}

With all the relevant operators at hand it is straightforward
to derive the NRG equation for the NLL  potential running of the
spin-flip potential. It reads
\bea
{d D^{(2)}_{S^2,s} \over d\ln\nu}&=&
-2 m_r^3 \left( {1\over m_1^3} +
{1\over m_2^3} \right)  C_F^2\alpha_{V_s}^2 D^{(2)}_{S^2,s}
+{m_r^2 \over m_1m_2}\,
C_F^2\alpha_{V_s}\Bigg(2D^{(2)}_{d,s}D^{(2)}_{S^2,s}
\nn\\
&&
+{8 \over 3}
\left(D^{(2)}_{S^2,s}\right)^2
-8 D^{(2)}_{S^2,s} D_{1,s}^{(2)}
-{5 \over 12} \left(D_{S_{12,s}}^{(2)}\right)^2  \Bigg)
-C_AC_FD_s^{(1)} D^{(2)}_{S^2,s}
\nn\\
&&+{1 \over 2}\alpha_s^3C_F^2 m_r^2
\left({c_{pp'}^{(1)}\over m_1^2}{c_F^{(2)}}
+{c_F^{(1)}}{c_{pp'}^{(2)}\over m_2^2}\right)
-2\alpha_s^3C_F^2 {m_r^2 \over m_1m_2}c_F^{(1)}c_F^{(2)}
\nn\\
&&
-{1 \over 4}\alpha_s^3C_F^2 {m_r^2 \over m_1m_2}c_S^{(1)}c_S^{(2)}
-\alpha_s^3C_F^2 m_r^2
\left({c_F^{(1)}}{c_S^{(2)}\over m_2^2}
+{c_S^{(1)}\over m_1^2}{c_F^{(2)}}\right)
\nn\\
&&
+{1 \over 2}\alpha_s^3C_F(4C_F-C_A) m_r
\left({c_F^{(1)}}{c_S^{(2)}\over m_2}+{c_S^{(1)}\over m_1}{c_F^{(2)}}\right)
\nn\\
&&
+{1 \over 8}\alpha_s^3C_FC_Am_r
\left({c_F^{(1)}}{c_F^{(2)\,2}\over m_2}+{c_F^{(1)\,2}\over m_1}{c_F^{(2)}}\right)
-{C_A^2(C_A-2C_F)\alpha_s^3 \over 2}D^{(2)}_{S^2,1/r^3}\,,
\label{Dsnllp}
\eea
where $m_r = m_1m_2/(m_1+m_2)$ is the reduced mass. The
notation for the Wilson coefficients is adopted from
Ref.~\cite{Pin2}, where also explicit results can be found.
The first  two lines of Eq.~(\ref{Dsnllp}) correspond to
the double insertion of the  ${\cal O}(v^2,\alpha_sv)$ operators
discussed in Appendix~A.3. Note that the contribution
proportional to $\left(D_{S_{12,s}}^{(2)}\right)^2$
disagrees with the result of Ref.~\cite{MSQED}.
To reproduce the ${\cal O}(\alpha^2)$ corrections to the positronium
hyperfine splitting \cite{Pachucki,CMY} in Ref.~\cite{MSQED} the factor  $9$
of the corresponding Wilson coefficient $U_t^2$ should be changed to  $5$.
This was corrected in Ref.~\cite{Manohar:2000cg}.
The expressions in the third and forth line of Eq.~(\ref{Dsnllp})
follow from the results for Appendix~A.1. The fifth and the first term of the
sixth line are obtained from Appendix~A.2. Finally, the last term in
Eq.~(\ref{Dsnllp}) corresponds to the ultrasoft-potential running discussed in
Appendix~A.4.


\section*{Appendix B: Analytical result for the NNLL contribution to $c_s$}

The analytical results for $A_i$ and $f_i(z)$ of Eq.~(\ref{Gammannll}) read
\begin{eqnarray}
  &&
  f_{1}(z) = z^{3 \beta_0 - 2 C_A}
  {}_3F_2\left({1, 3 - \frac{2 C_A}{\beta_0}, 3 - \frac{2 C_A}{\beta_0}};{ 4 -
\frac{2 C_A}{\beta_0}, 4 - \frac{2 C_A}{\beta_0}};
\frac{z^{\beta_0}}{2}\right)\,,\quad
  \nonumber\\
  &&
  f_{2}(z)= z^{3 \beta_0-2 C_A}
  {}_2F_1\left(3 - \frac{2 C_A}{\beta_0}, 1; 4 - \frac{2 C_A}{\beta_0};
\frac{z^{\beta_0}}{2}\right)\,,\quad
  f_{3}(z) = z^{2 \beta_0 - (25 C_A)/6}\,,\quad
  \nonumber\\
  &&
  f_{4}(z) = z^{2 \beta_0 - 4 C_A}\,,\quad
  f_{5}(z) = z^{2 \beta_0 - 3 C_A}\,,\quad
  f_{6}(z) = z^{2 \beta_0 - 2 C_A}\,,\quad
  f_{7}(z) = z^{2 \beta_0 - C_A}\,,\quad
  \nonumber\\
  &&
  f_{8}(z) = z^{\beta_0 - (13 C_A)/6}\,,\quad
  f_{9}(z) = z^{\beta_0 - 2 C_A}\,,\quad
  f_{10}(z) = z^{\beta_0}\,,\quad
  f_{11}(z) = \ln(z)\,,\quad
  \nonumber\\
  &&
  f_{12}(z) = \ln^2(z)\,,\quad
  f_{13}(z) = \ln\left(2-z^{\beta_0}\right)\,,\quad
  f_{14}(z) = z^{2\beta_0 - 2 C_A}\ln\left(2-z^{\beta_0}\right) \,,\quad
  \nonumber\\
  &&
  f_{15}(z) = z^{\beta_0}\ln\left(2-z^{\beta_0}\right) \,,
  \nonumber\\
  && f_{16}(z) = 2^{2C_A/\beta_0}z^{2 \beta_0-2C_A}
 {}_3F_2\left({ - \frac{2 C_A}{\beta_0}, 2 - \frac{2 C_A}{\beta_0}, 2 - \frac{2
C_A}{\beta_0}};{ 3 -
\frac{2 C_A}{\beta_0}, 3 - \frac{2 C_A}{\beta_0}}; \frac{z^{\beta_0}}{2}\right)
\,,\nonumber\\
  &&
  f_{17}(z) = 1\,,\quad
  f_{18}(z) = z^{2\beta_0}\,,\quad
  f_{19}(z) = \mbox{Li}_2\left(z^{\beta_0}/2\right)
  \,.
\end{eqnarray}

{\scalefont{0.85}
\begin{eqnarray}
A_{1} &=&
    \frac{C_F^3(-2 C_A^2 -6 C_A C_F-4 C_F^2)(C_A - 8 n_l T_F)}{
    3 (5 C_A - 4 n_l T_F) (9 C_A - 4 n_l T_F)^2 (2 C_A - n_l T_F)}
\,,\nonumber
\\
A_{2} &=&
-
    \frac{C_F^3(C_A^2+3 C_AC_F+2C_F^2) (C_A - 8 n_l T_F)}{
    4 (5 C_A - 4 n_l T_F) (9 C_A - 4 n_l T_F) (2 C_A - n_l T_F)^2 }
\,,\nonumber
\\
A_{3} &=&
-C_F^4n_l T_F
    \frac{(3840  C_A+6144  C_F)(-C_A + 8 n_l T_F)}{
    13 C_A(-5 C_A + 4 n_l T_F) (-9 C_A + 8 n_l T_F) (-19 C_A + 16 n_l T_F)^2}
\,,\nonumber
\\
A_{4} &=&
  -3C_F^4
    \frac{ (23 C_A - 4 n_l T_F) (C_A + 4 n_l T_F)}{
    8 (5 C_A - 4 n_l T_F)^4}
\,,\nonumber
\\
A_{5} &=&
    \frac{3 C_AC_F^3}{
    (13 C_A - 8 n_l T_F)^2}
\,,\nonumber
\\
A_{6} &=&
{C_F^2 \over \pi^2}
\Bigg[
C_F\Bigg(
    \frac{3 n_l T_F (1009 C_A^3 - 48 C_A^2 n_l T_F - 720 C_A n_l^2 T_F^2 + 256 n_l^3 T_F^3)}{
    32 (5 C_A - 4 n_l T_F) (11 C_A - 4 n_l T_F)^2 (2 C_A - n_l T_F)^2}
  \Bigg)
\nonumber\\&&
  +
    \frac{5975 C_A^5 - 37478 C_A^4 n_l T_F + 22592 C_A^3 n_l^2 T_F^2 + 18080 C_A^2 n_l^3 T_F^3 - 17408 C_A n_l^4 T_F^4 + 3584 n_l^5 T_F^5}{
    144 (5 C_A - 4 n_l T_F) (11 C_A - 4 n_l T_F)^2 (2 C_A - n_l T_F)^2}
        \Bigg]
\nonumber\\&&\mbox{}
+
  C_F^2\Bigg(
    \frac{C_A(C_A-6C_F)}{
    32 (2 C_A - n_l T_F)^2}
  \Bigg)
  +C_F^3\Bigg(
    \frac{-3 C_A (11 C_A - 16 n_l T_F)}{
    16 (5 C_A - 4 n_l T_F) (2 C_A - n_l T_F)^2}
  \Bigg)
\nonumber\\&&\mbox{}
  +C_F^4\Bigg(
    \frac{3 (1742 C_A^2 - 1697 C_A n_l T_F + 368 n_l^2 T_F^2)}{
    104 (5 C_A - 4 n_l T_F) (11 C_A - 4 n_l T_F) (2 C_A - n_l T_F)^2}
  \Bigg)
\nonumber\\&&\mbox{}
  +C_F^5\Bigg(
    \frac{3 (13 C_A - 32 n_l T_F) (C_A - 8 n_l T_F)}{
    104 C_A (5 C_A - 4 n_l T_F) (11 C_A - 4 n_l T_F) (2 C_A - n_l T_F)^2}
  \Bigg)
\,,\nonumber
\\
A_{7} &=& C_F^3\Bigg(
    \frac{12 C_A}{
    (19 C_A - 8 n_l T_F)^2}
  \Bigg)
  +C_F^4\Bigg(
    \frac{-36}{
    (19 C_A - 8 n_l T_F)^2}
  \Bigg)
\,,\nonumber
\\
A_{8} &=&
C_F^4n_l T_F\Bigg(
    \frac{34560 C_A +55296C_F}{
    13 (9 C_A - 8 n_l T_F)^3 (5 C_A - 4 n_l T_F)}
  \Bigg)
\,,\nonumber
\\
A_{9} &=&
{1 \over \pi^2}\Bigg[
  C_F^2\Bigg(
    \frac{-2 C_A (C_A - 8 n_l T_F) (397 C_A^2 - 385 C_A n_l T_F + 100 n_l^2 T_F^2)}{
    3 (5 C_A - 4 n_l T_F)^2 (11 C_A - 4 n_l T_F)^2}
  \Bigg)
\nonumber\\&&\mbox{}
  +C_F^3\Bigg(
    \frac{-2 (C_A - 8 n_l T_F) (121 C_A^2 - 70 C_A n_l T_F + 16 n_l^2 T_F^2)}{
    (5 C_A - 4 n_l T_F)^2 (11 C_A - 4 n_l T_F)^2}
  \Bigg)\Bigg]
\nonumber\\&&\mbox{}
+
  C_F^4\Bigg(
    \frac{36 C_A (501 C_A^3 + 706 C_A^2 n_l T_F - 1480 C_A n_l^2 T_F^2 + 448 n_l^3 T_F^3)}{
    (9 C_A - 8 n_l T_F) (5 C_A - 4 n_l T_F)^4 (11 C_A - 4 n_l T_F)}
  \Bigg)
\nonumber\\&&\mbox{}
  +C_F^5\Bigg(
    \frac{-72 (C_A - 8 n_l T_F) (3 C_A + 8 n_l T_F)}{
    (9 C_A - 8 n_l T_F) (5 C_A - 4 n_l T_F)^3 (11 C_A - 4 n_l T_F)}
  \Bigg)
\nonumber\\&&\mbox{}
+
{1 \over \pi^2}\ln\left(\frac{\nu_h^2}{m_q^2}\right)C_F^2\Bigg(
    \frac{ - 2 C_A (C_A - 8 n_l T_F)
        }{
    2 (5 C_A - 4 n_l T_F)^2 }
  \Bigg)
\,,\nonumber
\\
A_{10} &=&
-{C_F^2 \over \pi^2}\Bigg(
    \frac{C_A (35 C_A^2 - 164 C_A n_l T_F + 80 n_l^2 T_F^2+108C_Fn_lT_F)}{
    2 (5 C_A - 4 n_l T_F) (11 C_A - 4 n_l T_F)^2}
  \Bigg)
+
\nonumber\\&&\mbox{}
  +C_F^3\Bigg(
    \frac{108 C_A^2 (5 C_A + 4 n_l T_F)}{
    (5 C_A - 4 n_l T_F) (11 C_A - 4 n_l T_F)^3}
  \Bigg)
  +C_F^4\Bigg(
    \frac{108 C_A (533 C_A + 38 n_l T_F)}{
    13 (5 C_A - 4 n_l T_F) (11 C_A - 4 n_l T_F)^3}
  \Bigg)
\nonumber\\&&\mbox{}
  +C_F^5\Bigg(
    \frac{216 (221 C_A - 32 n_l T_F)}{
    13 (5 C_A - 4 n_l T_F) (11 C_A - 4 n_l T_F)^3}
  \Bigg)
\,,\nonumber
\\
A_{11} &=&
{1 \over \pi^2}\Bigg[
  C_F^3\Bigg(
    \frac{32 C_A^2 - 89 C_A n_l T_F + 32 n_l^2 T_F^2}{
    4 (5 C_A - 4 n_l T_F) (2 C_A - n_l T_F)}
  \Bigg)
\nonumber\\&&\mbox{}
  +C_F^2\Bigg(
    \frac{-43 C_A^3 - 120 C_A^2 T_F + 40 C_A^2 n_l T_F + 156 C_A n_l T_F^2
        - 16 C_A n_l^2 T_F^2 - 48 n_l^2 T_F^3}{
    6 (5 C_A - 4 n_l T_F) (2 C_A - n_l T_F)}
  \Bigg)\Bigg]
\nonumber\\&&\mbox{}
+
  C_F^2\Bigg(
    \frac{-C_A (C_A - 6 C_F)}{
    12 (2 C_A - n_l T_F)}
  \Bigg)
\nonumber\\&&\mbox{}
  +    \frac{C_F^3}{
    6 (13 C_A - 8 n_l T_F) (19 C_A - 8 n_l T_F) (5 C_A - 4 n_l T_F)
        (9 C_A - 4 n_l T_F) }
        \nonumber\\&&\mbox{}
        \times
        \frac{1}{
     (11 C_A - 4 n_l T_F)^2 (2 C_A - n_l T_F)}
        \Bigg(
        3 C_A (9 C_A - 4 n_l T_F) (263641 C_A^5 - 919114 C_A^4 n_l T_F
        \nonumber\\&&\mbox{}
        + 1071256 C_A^3 n_l^2 T_F^2 - 556448 C_A^2 n_l^3 T_F^3
        + 131456 C_A n_l^4 T_F^4 - 11264 n_l^5 T_F^5)
        \nonumber\\&&\mbox{}
        + 8 C_A^2 (C_A - 8 n_l T_F) (13 C_A - 8 n_l T_F) (19 C_A - 8 n_l T_F)
        \nonumber\\&&\mbox{}
        \times
        (11 C_A - 4 n_l T_F)^2
            {}_2F_1(1, 1, 4 - (2 C_A)/\beta_0, -1)
            \Bigg)
\nonumber\\&&\mbox{}
  +    \frac{C_F^4}{
    2 (19 C_A - 16 n_l T_F) (9 C_A - 8 n_l T_F)^2 (19 C_A - 8 n_l T_F)
        (5 C_A - 4 n_l T_F)^3 }
        \nonumber\\&&\mbox{}
        \times
        \frac{1}{
     (9 C_A - 4 n_l T_F) (11 C_A - 4 n_l T_F)^2 (2 C_A - n_l T_F)}
        \Bigg(
        -((9 C_A - 4 n_l T_F) (3565164294 C_A^9
        \nonumber\\&&\mbox{}
        - 13850749863 C_A^8 n_l T_F +
        20837783628 C_A^7 n_l^2 T_F^2 - 13367118064 C_A^6 n_l^3 T_F^3
        \nonumber\\&&\mbox{}
        - 187327680 C_A^5 n_l^4 T_F^4 + 5932724736 C_A^4 n_l^5 T_F^5
        - 3985068032 C_A^3 n_l^6 T_F^6
        \nonumber\\&&\mbox{}
        + 1224376320 C_A^2 n_l^7 T_F^7
        - 176160768 C_A n_l^8 T_F^8 + 8388608 n_l^9 T_F^9))
        \nonumber\\&&\mbox{}
        + 8 C_A (19 C_A - 16 n_l T_F) (C_A - 8 n_l T_F) (9 C_A - 8 n_l T_F)^2
        (19 C_A - 8 n_l T_F)
        \nonumber\\&&\mbox{}
        \times
        (5 C_A - 4 n_l T_F)^2 (11 C_A - 4 n_l T_F)^2
        {}_2F_1(1, 1, 4 - (2 C_A)/\beta_0, -1)
        \Bigg)
\nonumber\\&&\mbox{}
  +
    \frac{C_F^5}{
    3 (19 C_A - 16 n_l T_F) (9 C_A - 8 n_l T_F)^2
        (5 C_A - 4 n_l T_F)^2 (9 C_A - 4 n_l T_F) }
        \nonumber\\&&\mbox{}
        \times
        \frac{1}{
     (11 C_A - 4 n_l T_F)^2 (2 C_A - n_l T_F)}
        \Bigg(-3 (9 C_A - 4 n_l T_F) (9786501 C_A^6 - 23721912 C_A^5 n_l T_F
        \nonumber\\&&\mbox{}
        + 22193456 C_A^4 n_l^2 T_F^2 - 11489920 C_A^3 n_l^3 T_F^3
        + 4389888 C_A^2 n_l^4 T_F^4
        \nonumber\\&&\mbox{}
        - 1171456 C_A n_l^5 T_F^5
        + 131072 n_l^6 T_F^6)
        \nonumber\\&&\mbox{}
        + 8 (19 C_A - 16 n_l T_F) (C_A - 8 n_l T_F)
        (9 C_A - 8 n_l T_F)^2
        \nonumber\\&&\mbox{}
        \times
        (5 C_A - 4 n_l T_F) (11 C_A - 4 n_l T_F)^2
        {}_2F_1(1, 1, 4 - (2 C_A)/\beta_0, -1)
        \Bigg)
\nonumber\\
 &&
+\ln(2)\Bigg[
{2 T_FC_F^2 \over \pi^2}
  +C_F^3\Bigg(
    \frac{576 C_A^3}{
    (5 C_A - 4 n_l T_F) (11 C_A - 4 n_l T_F)^2}
  \Bigg)
\nonumber\\&&\mbox{}
  +C_F^4\Bigg(
    \frac{1728 C_A^2}{
    (5 C_A - 4 n_l T_F) (11 C_A - 4 n_l T_F)^2}
  \Bigg)
  +C_F^5\Bigg(
    \frac{1152 C_A}{
    (5 C_A - 4 n_l T_F) (11 C_A - 4 n_l T_F)^2}
  \Bigg)
\Bigg]
\nonumber\\
 &&
+B_{1/2}\left(2-\frac{2C_A}{\beta_0},1+\frac{2C_A}{\beta_0}\right)
\left[\frac{8C_AC_F^2(C_A-2C_F)}{(11 C_A - 4 n_l
T_F)}\right]
\nonumber\\
 &&
+{1 \over
\pi^2}\ln\left(\frac{\nu_h^2}{m_q^2}\right)C_F^2\Bigg(
    \frac{ - C_A (5 C_A - 4 n_l T_F) (11 C_A - 4 n_l T_F) }{
    2 (5 C_A - 4 n_l T_F)^2 }
  \Bigg)
\,,\nonumber
\\
A_{12} &=&
C_F^4\Bigg(
    \frac{-72 C_A^2 (48 C_A^2 - 59 C_A n_l T_F + 16 n_l^2 T_F^2)
        -36 C_A C_F(3 C_A + 8 n_l T_F)(5 C_A - 4 n_l T_F)}{
    (9 C_A - 8 n_l T_F) (5 C_A - 4 n_l T_F)^2 (11 C_A - 4 n_l T_F)}
  \Bigg)
\,,\nonumber
\\
A_{13} &=&
  C_F^3\Bigg(
    \frac{(3456(C_A^2+2C_F^2)+10368 C_AC_F)C_A (2 C_A - n_l T_F)^2 }{
    2 (5 C_A - 4 n_l T_F) (11 C_A - 4 n_l T_F)^3 (2 C_A - n_l T_F)^2 }
  \Bigg)
\,,\nonumber
\\
A_{14} &=&
-C_F^3\Bigg(
    \frac{ ( 3 C_A^2+ 9 C_AC_F+6C_F^2) (C_A - 8 n_l T_F) (11 C_A - 4 n_l T_F)^2
        }{
    2 (5 C_A - 4 n_l T_F) (11 C_A - 4 n_l T_F)^3 (2 C_A - n_l T_F)^2 }
  \Bigg)
\,,\nonumber
  \\
A_{15} &=&
-C_F^3\Bigg(
    \frac{ ( 1728 C_A^2+ 5184 C_AC_F+ 3456 C_F^2)C_A (2 C_A - n_l T_F)^2
        }{
    2 (5 C_A - 4 n_l T_F) (11 C_A - 4 n_l T_F)^3 (2 C_A - n_l T_F)^2 }
  \Bigg)
\,,\nonumber
  \\
A_{16} &=&
    \frac{ - 3 C_A C_F^2(C_A-2C_F)}{
    32(2 C_A -  n_l T_F)^2}
\,,\nonumber
  \\
A_{17} &=&
{1 \over \pi^2}\Bigg[
  C_F^2\Bigg(
    \frac{1}{
    144 (5 C_A - 4 n_l T_F)^2 (11 C_A - 4 n_l T_F)^2 (2 C_A - n_l T_F)^2}
        \nn\\&&\mbox{}
        \times
        (
        172973 C_A^6 - 1635462 C_A^5 n_l T_F + 2956992 C_A^4 n_l^2 T_F^2
        - 2335616 C_A^3 n_l^3 T_F^3
        \nonumber\\&&\mbox{}
        + 940032 C_A^2 n_l^4 T_F^4
        - 187392 C_A n_l^5 T_F^5 + 14336 n_l^6 T_F^6)
  \Bigg)
\nonumber\\&&\mbox{}
  +C_F^3\Bigg(
    \frac{1}{
    32 (5 C_A - 4 n_l T_F)^2 (11 C_A - 4 n_l T_F)^2 (2 C_A - n_l T_F)^2}
        \nn\\&&\mbox{}
        \times
        (
        30976 C_A^5 - 277279 C_A^4 n_l T_F + 371548 C_A^3 n_l^2 T_F^2
        - 200144 C_A^2 n_l^3 T_F^3
        \nonumber\\&&\mbox{}
        + 50240 C_A n_l^4 T_F^4 - 5120 n_l^5 T_F^5
        )
  \Bigg)\Bigg]
  +C_F^2\Bigg(
    \frac{-C_A (C_A - 6 C_F))}{
    32(2 C_A - n_l T_F)^2}
  \Bigg)
\nonumber\\&&\mbox{}
  -\Bigg(
    \frac{3 C_AC_F^3}{
    16 (13 C_A - 8 n_l T_F)^2 (19 C_A - 8 n_l T_F)^2 (5 C_A - 4 n_l T_F)
        (11 C_A - 4 n_l T_F)^3 (2 C_A - n_l T_F)^2}
        \nonumber\\&&\mbox{}
        \times
        (251269951 C_A^8 + 767113748 C_A^7 n_l T_F
        - 3841834880 C_A^6 n_l^2 T_F^2 + 5696836928 C_A^5 n_l^3 T_F^3
        \nonumber\\&&\mbox{}
        - 4303456256 C_A^4 n_l^4 T_F^4 + 1851312128 C_A^3 n_l^5 T_F^5
        - 454406144 C_A^2 n_l^6 T_F^6
        \nonumber\\&&\mbox{}
        + 58130432 C_A n_l^7 T_F^7 - 2883584 n_l^8 T_F^8)
  \Bigg)
\nonumber\\&&\mbox{}
  -\Bigg(
    \frac{3 C_F^4}{
    8 (19 C_A - 16 n_l T_F)^2 (9 C_A - 8 n_l T_F)^3 (19 C_A - 8 n_l T_F)^2
        (5 C_A - 4 n_l T_F)^4 }
        \nonumber\\&&\mbox{}
        \frac{1}{
     (11 C_A - 4 n_l T_F)^3 (2 C_A - n_l T_F)^2}
        (900867787121646 C_A^{14} - 6198841921859001 C_A^{13} n_l T_F
        \nonumber\\&&\mbox{}
        + 19119275729072316 C_A^{12} n_l^2 T_F^2 - 34644247240650784 C_A^{11} n_l^3 T_F^3
        \nonumber\\&&\mbox{}
        + 40320030084122112 C_A^{10} n_l^4 T_F^4
        - 30355280730121984 C_A^9 n_l^5 T_F^5
        \nonumber\\&&\mbox{}
        + 13082614024074240 C_A^8 n_l^6 T_F^6 - 558681493807104 C_A^7 n_l^7 T_F^7
        \nonumber\\&&\mbox{}
        - 3412540371369984 C_A^6 n_l^8 T_F^8 + 2548850603065344 C_A^5 n_l^9 T_F^9
        \nonumber\\&&\mbox{}
        - 1019296780124160 C_A^4 n_l^{10} T_F^{10}
        + 251825139744768 C_A^3 n_l^{11} T_F^{11}
        \nonumber\\&&\mbox{}
        - 37363262685184 C_A^2 n_l^{12} T_F^{12}
        + 2860448219136 C_A n_l^{13} T_F^{13}
        - 68719476736 n_l^{14} T_F^{14})
  \Bigg)
\nonumber\\&&\mbox{}
  -\Bigg(
    \frac{3 C_F^5}{
    8 (19 C_A - 16 n_l T_F)^2 (9 C_A - 8 n_l T_F)^3 (5 C_A - 4 n_l T_F)^3
        (11 C_A - 4 n_l T_F)^3 (2 C_A - n_l T_F)^2}
        \nonumber\\&&\mbox{}
        \times
        (250339248081 C_A^{10} - 1197788138208 C_A^9 n_l T_F
        + 2595380074848 C_A^8 n_l^2 T_F^2
        \nonumber\\&&\mbox{}
        - 3536482592256 C_A^7 n_l^3 T_F^3
        + 3565032448256 C_A^6 n_l^4 T_F^4 - 2818753511424 C_A^5 n_l^5 T_F^5
        \nonumber\\&&\mbox{}
        + 1679893954560 C_A^4 n_l^6 T_F^6
        - 692836106240 C_A^3 n_l^7 T_F^7
        + 178875531264 C_A^2 n_l^8 T_F^8
        \nonumber\\&&\mbox{}
        - 24964497408 C_A n_l^9 T_F^9
        + 1342177280 n_l^{10} T_F^{10})
  \Bigg)
\nonumber\\&&\mbox{}
+\pi^2\Bigg[
 C_F^3\Bigg(
    \frac{144 C_A^3}{
    (5 C_A - 4 n_l T_F) (11 C_A - 4 n_l T_F)^3}
  \Bigg)
\nonumber\\&&\mbox{}
  +C_F^4\Bigg(
    \frac{432 C_A^2}{
    (5 C_A - 4 n_l T_F) (11 C_A - 4 n_l T_F)^3}
  \Bigg)
  +C_F^5\Bigg(
    \frac{288 C_A}{
    (5 C_A - 4 n_l T_F) (11 C_A - 4 n_l T_F)^3}
  \Bigg)
\Bigg]
\nonumber\\
 &&
+\ln^2(2)\Bigg(
\frac{-864 C_A^3C_F^3-2592 C_A^2C_F^4-1728 C_AC_F^5}{
    (5 C_A - 4 n_l T_F) (11 C_A - 4 n_l T_F)^3}
\Bigg)
\nonumber\\
 &&
+{}_2F_1\left(1,1;4-\frac{2C_A}{\beta_0};-1\right)\Bigg[
  C_F^3\Bigg(
    \frac{C_A^2 (C_A - 8 n_l T_F)}{
    2 (5 C_A - 4 n_l T_F) (9 C_A - 4 n_l T_F) (2 C_A - n_l T_F)^2}
  \Bigg)
\nonumber\\&&\mbox{}
  +C_F^4\Bigg(
    \frac{3 C_A (C_A - 8 n_l T_F)}{
    2 (5 C_A - 4 n_l T_F) (9 C_A - 4 n_l T_F) (2 C_A - n_l T_F)^2}
  \Bigg)
\nonumber\\&&\mbox{}
  +C_F^5\Bigg(
    \frac{C_A - 8 n_l T_F}{
    (5 C_A - 4 n_l T_F) (9 C_A - 4 n_l T_F) (2 C_A - n_l T_F)^2}
  \Bigg)
\Bigg]
\nonumber\\
 &&
+{}_3F_2\left(1,3-\frac{2C_A}{\beta_0},3-\frac{2C_A}{\beta_0}
;4-\frac{2C_A}{\beta_0},4-\frac{2C_A}{\beta_0}; \frac{1}{2}\right)
\nn\\&&\mbox{}
\times
\Bigg[
  C_F^3\Bigg(
    \frac{2 C_A^2 (C_A - 8 n_l T_F)}{
    3 (5 C_A - 4 n_l T_F) (9 C_A - 4 n_l T_F)^2 (2 C_A - n_l T_F)}
  \Bigg)
\nonumber\\&&\mbox{}
  +C_F^4\Bigg(
    \frac{2 C_A (C_A - 8 n_l T_F)}{
    (5 C_A - 4 n_l T_F) (9 C_A - 4 n_l T_F)^2 (2 C_A - n_l T_F)}
  \Bigg)
\nonumber\\&&\mbox{}
  +C_F^5\Bigg(
    \frac{4 (C_A - 8 n_l T_F)}{
    3 (5 C_A - 4 n_l T_F) (9 C_A - 4 n_l T_F)^2 (2 C_A - n_l T_F)}
  \Bigg)\Bigg]
\nonumber\\
&&\mbox{}
+2^{2C_A/\beta_0}{}_3F_2\left({ - \frac{2 C_A}{\beta_0}, 2 - \frac{2
C_A}{\beta_0}, 2 - \frac{2
C_A}{\beta_0}};{ 3 -
\frac{2 C_A}{\beta_0}, 3 - \frac{2 C_A}{\beta_0}};
\frac{1}{2}\right)\left[
 \frac{  3 C_A C_F^2(C_A-2C_F)}{
    32(2 C_A -  n_l T_F)^2}\right]
\nonumber\\
&&\mbox{}
+
{1 \over \pi^2}\ln\left(\frac{\nu_h^2}{m_q^2}\right)
C_F^2\Bigg(
    \frac{ C_A (C_A - 8 n_l T_F)  }{
     (5 C_A - 4 n_l T_F)^2 }
  \Bigg)
\,,\nonumber
\\
A_{18} &=&
    \frac{-3C_F^4}{
    8 (11 C_A - 4 n_l T_F)^2}
\,,\nonumber
\\
A_{19} &=&
    \frac{-1728 C_A^3C_F^3-5184 C_A^2C_F^4-3456 C_AC_F^5}{
    (5 C_A - 4 n_l T_F) (11 C_A - 4 n_l T_F)^3}
\,,
\end{eqnarray}
}
where $B_z$ is the
incomplete beta-function,
and ${}_pF_q$ is the
hypergeometric function.

\section*{Appendix C:
$O(\alpha^4\ln^2\alpha)$ and $O(\alpha^5\ln^3\alpha)$ corrections to the
ratio of ortho- and parapositronium  decay rates}

We can also obtain novel results for the ratio of the ortho- ($oPs$)  and
parapositronium ($pPs$) decay rates by considering the Abelian
limit of the previous computation and taking into account the modifications of
the NRQCD matching coefficients due to the annihilation terms.
The series for the decay rates ratio reads
\bea
{\Gamma(oPs\to3\gamma) \over \Gamma(pPs\to2\gamma)}
&=&
{4(\pi^2-9) \over 9\pi}\alpha
\left\{
1+\left(5-{\pi^2\over 4}+A_o\right){\alpha\over\pi}
+{7\over 3}\alpha^2\ln\alpha
\right.
\nn\\
&&
+\left[\left(5-{\pi^2 \over 4}\right)^2+\left(5-{\pi^2\over 4}\right)A_o+B_o-B_p
\right]\left({\alpha\over\pi}\right)^2
\nn\\
&&
-\left[
-\frac{73}{9 } - \frac{7\, {A_o}}{3 } + \frac{7\,\pi^2 }{12} + {2\,\log (2)}
\right]
{\alpha^3 \over \pi}\ln\alpha
\nn\\
&&
\left.
+{83 \over 36}\alpha^4\ln^2\alpha-{7 \over 6\pi}{\alpha^5}\ln^3\alpha
+\cdots
\right\}\,,
\eea
where $A_o=10.286606(10)$, $B_o=44.87(26)$ and $B_p=5.1243(33)$.
The first four correction terms in the curly brackets
can be extracted from know results (see {\it e.g.} \cite{Pen}).
The last two terms are our new results.




\begin{thebibliography}{99}

\bibitem{AppPol} T. Appelquist and H.D. Politzer,
Phys.\ Rev.\ Lett.\ {34} (1975)  43.

\bibitem{CasLep} W.E. Caswell and G.P. Lepage,
Phys.\ Lett.\ B {167} (1986) 437.

\bibitem{BBL} G.T. Bodwin, E. Braaten, and G.P. Lepage,
Phys.\ Rev.\ D {51} (1995) 1125; Erratum {\it ibid.}\ {55} (1997) 5853.

\bibitem{Hag} K. Hagiwara {\it et al.},
Phys.\ Rev.\ D {66} (2002) 010001.

\bibitem{MarMiq} M. Martinez and  R. Miquel,
Eur.\ Phys.\ J.\ C {27} (2003) 49.

\bibitem{KMRR} W. Kwong, P.B. Mackenzie, R. Rosenfeld, and J.L. Rosner,
Phys.\ Rev.\ D {37} (1988) 3210, and references therein.

\bibitem{KPP} J.H. K\"uhn, A.A. Penin, and A.A. Pivovarov,
Nucl.\ Phys.\ {B 534} (1998) 356.

\bibitem{CzaMel1} A. Czarnecki and K. Melnikov,
Phys.\ Rev.\ Lett.\ {80} (1998) 2531.

\bibitem{BSS} M. Beneke, A. Signer, and V.A. Smirnov,
Phys.\ Rev.\ Lett.\ {80} (1998) 2535.

\bibitem{HoaTeu} A.H. Hoang and  T. Teubner,
Phys.\ Rev.\ D {58} (1998) 114023.

\bibitem{PenPiv1} A.A. Penin and A.A. Pivovarov,
Phys.\ Lett.\ B {435} (1998) 413;
Nucl.\ Phys.\ {B 549} (1999) 217.

\bibitem{MelYel1} K. Melnikov and A. Yelkhovsky,
Phys.\ Rev.\ D {59} (1999) 114009.

\bibitem{PenPiv2} A.A. Penin and A.A. Pivovarov,
Nucl.\ Phys.\ {B 550} (1999) 375;
Yad.\ Fiz.\ {64} (2001) 323
[Phys.\ Atom.\ Nucl.\ {64} (2001) 275].

\bibitem{CzaMel2} A. Czarnecki and K. Melnikov,
Phys.\ Rev.\ D {65} (2002) 051501;
Phys.\ Lett.\ B {519} (2001) 212.

\bibitem{HMST} A.H. Hoang, A.V. Manohar, I.W. Stewart, and T. Teubner,
Phys.\ Rev.\ Lett.\ {86}  (2001) 1951.

\bibitem{Pin2} A. Pineda,
Phys.\ Rev.\ D {65} (2002) 074007; Phys.\ Rev.\ D
{66} (2002) 054022.

\bibitem{HoaSte} A.H. Hoang and I.W. Stewart,
Phys.\ Rev.\ D {67} (2003) 114020.

\bibitem{Hoa} A.H. Hoang, Phys.\ Rev.\ D {69} (2004) 034009.

\bibitem{KalSar}
G. K{\"a}llen and A. Sarby,
K.\ Dan.\ Vidensk.\ Selsk.\ Mat.-Fis.\ Medd.\ {29},  N17 (1955) 1.

\bibitem{HarBro} I. Harris and L.M. Brown, Phys.\ Rev.\ {105} (1957) 1656.

\bibitem{PinSot1} A. Pineda and J. Soto,
Nucl.\ Phys.\ Proc.\ Suppl. {64}  (1998) 428.

\bibitem{KniPen1} B.A. Kniehl and A.A. Penin,
Nucl.\ Phys.\ {B 563} (1999) 200.

\bibitem{BPSV2} N. Brambilla, A. Pineda, J. Soto, and A. Vairo,
Nucl.\ Phys.\ {B 566} (2000) 275.

\bibitem{PinSot2} A. Pineda and J. Soto,
Phys.\ Lett.\ B {420} (1998) 391;
Phys.\ Rev.\ D {59} (1999) 016005.

\bibitem{CMY} A. Czarnecki, K. Melnikov, and A. Yelkhovsky,
Phys.\ Rev.\ A {59}  (1999) 4316.

\bibitem{Beneke:1999qg}
M.~Beneke, A.~Signer, and V.~A.~Smirnov,
Phys.\ Lett.\ B {454} (1999) 137.

\bibitem{KPSS1} B.A. Kniehl, A.A. Penin, V.A. Smirnov, and M. Steinhauser,
Phys.\ Rev.\ D {65} (2002) 091503;
Nucl.\ Phys.\ {B 635} (2002) 357.

\bibitem{BenSmi} M. Beneke and V.A. Smirnov,
Nucl.\ Phys.\ {B 522} (1998) 321.

\bibitem{Smi} V.A. Smirnov,
{\it Applied Asymptotic Expansions in Momenta and Masses}
(Springer-Verlag, Heidelberg, 2001).

\bibitem{KPSS2} B.A. Kniehl, A.A. Penin, V.A. Smirnov, and M. Steinhauser,
Phys.\ Rev.\ Lett.\ {90}  (2003) 212001;
Erratum {\it ibid.}\ {91} (2003) 139903.

\bibitem{LMR} M.E. Luke, A.V. Manohar, and I.Z. Rothstein,
Phys.\ Rev.\ D {61} (2000) 074025.

\bibitem{KniPen2} B.A. Kniehl and A.A. Penin,
Nucl.\ Phys.\ {B 577} (2000) 197.

\bibitem{KniPen3} B.A. Kniehl and A.A. Penin,
Phys.\ Rev.\ Lett.\ {85} (2000) 1210;
Erratum {\it ibid.}\ {85}  (2000) 3065;
Phys.\ Rev.\ Lett.\ {85} (2000) 5094.

\bibitem{HilLep} R.J. Hill and G.P. Lepage,
Phys.\ Rev.\ D {62}  (2000) 111301.

\bibitem{MelYel2} K. Melnikov and A. Yelkhovsky,
Phys.\ Rev.\ D {62}  (2000) 116003.

\bibitem{KPPSS} B.A. Kniehl, A.A. Penin, A. Pineda,
V.A. Smirnov, and M. Steinhauser,
DESY-03-172, TTP-03-40, UB-ECM-PF-03-28, hep-ph/0312086,
to appear in Phys.\ Rev.\ Lett.\ 

\bibitem{PPSS} A.A. Penin, A. Pineda, V.A. Smirnov, and M. Steinhauser,
DESY 04-042, TTP04-06, UB-ECM-PF-04-05, hep-ph/0403080,
to appear in Phys.\ Lett.\ B.

\bibitem{MS} A.~V.~Manohar and I.~W.~Stewart,
Phys.\ Rev.\ D {\bf 62}, 014033 (2000).

\bibitem{PenSte} A.A. Penin and M. Steinhauser,
Phys.\ Lett.\ B {538} (2002) 335.

\bibitem{Vol} M.B. Voloshin,
Nucl.\ Phys.\ {B 154}  (1979) 365;
Yad.\ Fiz.\ {36}  (1982) 247
[Sov.\ J. Nucl.\ Phys.\ {36}  (1982) 143].

\bibitem{Leu} H. Leutwyler,
Phys.\ Lett.\ B {98}  (1981) 447.

\bibitem{TitYnd2} S. Titard and F.J. Yndur\'ain,
Phys.\ Rev.\ D {51}  (1995) 6348.

\bibitem{Pin1} A. Pineda, Nucl.\ Phys.\ {B 494} (1997) 213.

\bibitem{Pin3} A. Pineda, Acta Phys.\ Polon.\ B {34} (2003) 5295.

\bibitem{BC} E. Braaten and Y.Q. Chen,
Phys.\ Rev.\ D {57} (1998) (1998) 4236; Erratum {\it ibid.}\ D {59} (1999) 079901.

\bibitem{sum} Y. Sumino, Phys.\ Rev.\ D {65} (2002) 054003; S. Recksiegel and
Y. Sumino Phys.\ Rev.\ D 65 (2002) 054018; A. Pineda, J.\ Phys.\ G 29 (2003) 371;
T. Lee, Phys.\ Rev.\ D 67 (2003) 014020.

\bibitem{Man} A.V. Manohar,
Phys.\ Rev.\ D {56} (1997) 230.

\bibitem{BPSV1} N. Brambilla, A. Pineda, J. Soto, and A. Vairo,
Phys.\ Lett.\ B {470} (1999) 215.

\bibitem{MSQED} A.V. Manohar and I.W. Stewart,
Phys.\ Rev.\ Lett.\ {85}  (2000) 2248.

\bibitem{Pachucki} K. Pachucki, Phys.\ Rev.\  A {56}  (1997) 297.

\bibitem{Manohar:2000cg}
A.~V.~Manohar and I.~W.~Stewart,
Nucl.\ Phys.\ Proc.\ Suppl.\  {94} (2001) 130.

\bibitem{Pen} A.A. Penin, hep-ph/0308204, and references therein.

\end{thebibliography}
\end{document}